\documentclass[a4paper,11pt]{article}
\usepackage{jheppub}
\usepackage{epsfig}
\usepackage[T1]{fontenc} 
\usepackage{rotating}

\usepackage{hyperref}
\usepackage{amssymb,amsmath}
\usepackage[caption=false]{subfig}

\usepackage{breqn}

\newcommand\eps{\epsilon}

\newcommand\cO{  {\cal O}  }
\newcommand{\bea}{\begin{eqnarray}}
\newcommand{\eea}{\end{eqnarray}}

\newcommand\bu{{\beta}_u}
\newcommand\bv{{\beta}_v}
\newcommand\buv{{\beta}_{uv}}

\newcommand\Li{{\rm Li}}

\def\be{\begin{equation}}
\def\ee{\end{equation}}

\allowdisplaybreaks[4]

\title{
Lectures on differential equations for Feynman integrals
}

\author[a]{Johannes M.\ Henn}
\affiliation[a]{Institute for Advanced Study, Princeton, NJ 08540, USA}
\emailAdd {jmhenn@ias.edu}

\abstract{
Over the last year significant progress was made in the understanding
of the computation of Feynman integrals using differential equations.
These lectures give a review of these developments, while not
assuming any prior knowledge of the subject.
After an introduction to differential equations for Feynman integrals,
we point out how they can be simplified using algorithms available in the mathematical
literature. We discuss how this is related to a recent conjecture for a canonical
form of the equations.
We also discuss a complementary approach that is based on 
properties of the space-time loop integrands,
and explain how the ideas of leading singularities and d-log representations
can be used to find an optimal basis for the differential equations.
Finally, as an application of these ideas we show how 
single-scale integrals can be bootstrapped using the Drinfeld associator
of a differential equation.
}


\begin{document}

\maketitle
\flushbottom

\newpage

\section{Introduction}

Feynman integrals are ubiquitous in quantum field theory.
They occur when quantities such as correlation functions of local operators,
or scattering amplitudes are computed in perturbation theory.
Beyond the lowest order in perturbation theory, 
integrals over $D$-dimensional space time
over (rational) propagator factors have to be evaluated.

Our ability to evaluate Feynman integrals in perturbation theory is
critical for connecting quantum field theory to experiment.
For example, in order to make precise theoretical predictions for collisions 
at the Large Hadron Collider (LHC), higher-order Feynman integrals
for complicated kinematical processes are a crucial ingredient, see e.g. \cite{Gehrmann:2014vva}. 
This not only applies to virtual amplitudes, but, via the optical theorem,
also to phase space integrals and to cross sections \cite{Anastasiou:2002yz,Anastasiou:2003yy}.
Other applications include statistical field theory, where critical exponents
can be computed; higher order calculations are also important for instance
for a better theoretical description of the anomalous magnetic moments.

{}From the mathematical point of view it is an interesting question to
ask what class of functions Feynman integrals give rise to.
These are typically functions of a set of parameters, such as the particle
momenta and masses in the case of scattering amplitudes, or a set of points 
in the case of correlation functions.
Experience shows that typical functions occurring in Feynman integrals
are certain classes of iterated integrals, elliptic functions, and possibly generalizations thereof.
It is an interesting open problem in general to predict, for a given Feynman graph,
what class of functions it is described by.
Understanding the above questions also implies, for fixed values of the
external parameters, an understanding of what periods Feynman integrals give rise to.
Periods are a class of numbers that is situated between the rational and irrational numbers
\cite{periods}.

Some Feynman integrals have divergences that can come from
different regions of integration, either ultraviolet (short distances) or infrared (long distances).
In that case, integrals can be defined in $D=4-2\epsilon$ dimensions, and
one is usually interested in the solution as a Laurent series in $\eps$.
Physically relevant quantities can be defined via a renormalization procedure,
such that divergences ultimately cancel, but it is useful to be able to compute
the $\eps$ expansion, in principle to any order in $\eps$.
From the mathematical point of view, the question about periods and functions can
be posed in a well-defined way for each term in the $\eps$ expansion.

In these lectures we present the method of differential equations (DE) for computing
Feynman integrals. It has a long and successful history \cite{Kotikov:1990kg,Remiddi:1997ny,Gehrmann:1999as}.
Earlier reviews include \cite{Argeri:2007up}.
Despite the success of the method, so far it was mostly used on a case by case basis,
with considerable amount of work needed by hand to solve the equations.
Over the last year a more systematic picture has emerged, and it is the goal of these lecture notes to present this in a coherent fashion.

One of the key new ideas of ref. \cite{Henn:2013pwa} is to choose an optimal basis
of integrals that leads to a system of differential equations in a canonical form.
We present two different, complimentary approaches to this end,
one of which is more algebraic, 
the other one more geometric in nature.

Functions coming from Feynman integrals have regular singularities as kinematic
variables approach a given singular point.
At the level of the differential equations, the canonical form should make the 
behavior of the Feynman integrals near singular points manifest.
This can be done using known algorithms available in the mathematical literature.
Having put the kinematic dependence into a canonical form, 
the next step is to simplify the dependence on the dimensional regularization parameter $\eps$. 
For example, in cases where the answer is given by iterated integrals, the idea
is to make this iterative structure manifest at the level of the differential equations.
We review the features of this algebraic approach, and explain how elliptic and more
complicated functions arise  from this viewpoint.

The approach mentioned above is an algebraic one, which to a large extent could be
applied to any system having regular singularities only. 
There is a complimentary approach that allows to gain insights on an appropriate
basis choice already at the level of the Feynman integrand.
The idea is to analyze the structure of (generalized) unitarity cuts
of the latter \cite{bookEden,Cachazo:2008vp,ArkaniHamed:2010kv,ArkaniHamed:2010gh},
where, roughly speaking, a number of propagators is replaced by delta functions.
It had been observed earlier that the analysis of generalized cuts (and in particular, leading singularities) is a natural way to 
choose a basis for loop integrands that make expressions for scattering amplitudes simple, 
and e.g. exhibit their properties in soft limits \cite{ArkaniHamed:2010kv,Drummond:2010mb}.
In ref. \cite{Henn:2013pwa} it was proposed to use these ideas in the context of differential equations in order to simplify the latter.
In that context, considering cuts of integrals is very natural,
since the cut integrals satisfy the same differential equations \cite{Anastasiou:2003yy}, 
albeit with different boundary conditions. Moreover, the cuts can be viewed as projection operators
onto sectors of the DE, which is a useful feature in practice.
This explains why cuts are an efficient guide to
finding a canonical form of the DE.

A closely related tool are certain d-log representations, either at the
level of parameter integrals, or directly at the level of loop integrals.
In the case where the answer is given by iterated integrals, the former can be 
used to prove transcendental weight properties of the answer,
and guarantee that a canonical form of the DE can be found.
In the latter case, the transcendental weight property is expected
conjecturally \cite{ArkaniHamed:2010gh}.

Over the last year, these tools have been applied successfully to the calculation of many classes
of Feynman integrals. In practice, often the most efficient strategy is to choose the basis
integrals according to their cut properties, which is most natural for Feynman integrals.
This usually leads to a form of the differential equations that is very close to the desired
canonical form, and the remaining algebraic transformations are then easily found.

These lecture notes are intended for students and researchers alike.
We hope that they are useful to phenomenologists desiring to get acquainted
with this method, in order to apply it to loop integrals relevant for their calculations, 
as well as to more formal theorists seeking a general understanding of multi-loop 
scattering amplitudes, and last but not least to mathematicians interested in new ideas 
that do not rely on the usual Feynman parametrization approach.

The style we have tried to adopt is one where the main ideas are explained with
the help of simple examples. In this way, we avoid the burden of a general but possibly
unintelligible notation, and are able to keep the lectures at a digestible length\footnote{
Cette amplification, que l'on confond si souvent avec le bien \'{e}crire, je la supporte de moins en moins. Quelle absurde n\'{e}cessit\'{e} de faire un article ou un livre ! O\`{u} trois lignes suffisent, je n'en mettrai pas une de plus. (A. Gide, Notes de journal, 1932)}. 
We frequently summarize general conclusions.

These lecture notes are organized as follows.
In section \ref{section_intro}, we recall basic definitions of Feynman integrals
and mention important properties that follow from them.
We continue in section \ref{section_de} to introduce notions that allow
to understand algebraic relations between different Feynman integrals
and to derive differential equations for them.
In section \ref{section_algorithm} we show how general properties of Feynman integrals
allow to transform the differential equations into a canonical form.
In section \ref{section_canonical} we discuss solutions in the case where the answer is given by iterated integrals.
In section \ref{section_leading_singularities}, we explain the analysis of generalized cuts / leading singularities
of Feynman integrals, and d-log representations to choose an optimal basis.
As an example, we fully explain the basis choice made in ref.  \cite{Henn:2013pwa} using these concepts.
The final section \ref{section_drinfeld}, which can be read independently, combines the ideas of the previous sections for a sample application, namely the computation of single-scale integrals
via differential equations.

\section{Definition and basic properties of Feynman integrals}
\label{section_intro}
Let us begin by recalling the main definitions and introduce the notation that we will use in the following. Our aim will be to be brief, as more details can be found in standard textbooks,
e.g. \cite{Smirnov:2004ym,bookPeskin}.
The main conclusions of this section are summarized at the end for the benefit of the reader
already familiar with this material.

\subsection{Definitions and Feynman parametrization}

We will discuss Feynman integrals in $D$-dimensional quantum
field theory. In the momentum space language, we have integrals over
$D$-dimensional space $d^{D}k$, with the integrand consisting of propagator factors 
like $1/[-(k+p)^2+m^2-i0]$.
The Feynman $i0$ prescription allows one to perform a Wick rotation 
from Minkowski space with metric $+-...-$ to Euclidean space, see e.g. 
\cite{Smirnov:2004ym,bookPeskin}.
In most of the following we drop the $i0$ from our formulas for simplicity of notation.

As an example, let us start with a momentum-space box integral at one loop.
This example will recur frequently in these notes.
\begin{align}
I_{\rm box} = \int \frac{d^{D}k}{i \pi^{D/2}} \frac{1}{k^2 (k+p_{1})^2 (k+p_1 +p_2)^2 (k-p_4)^2 }\,,
\end{align}
Here $p_{1},p_{2},p_{3},p_{4}$ are $D$-dimensional momenta  satisfying
momentum conservation ${\sum_{i=1}^{4} p_{i} = 0}$ and the on-shell conditions $p_{i}^2=0$.

Poincar\'{e} invariance implies that the integral depends on the Mandelstam invariants $s=(p_{1}+p_{2})^2$ and $t=(p_{2}+p_{3})^2$ only.
Moreover, the integral is covariant under dilatations, so that after normalizing it with some appropriate power of $s$ or $t$ it becomes a function of the dimensionless variable $x=t/s$ only.  

A very convenient notation for planar integrals as the above are dual (or region) coordinates.
This constitutes a change of variables.
We can rewrite the above integral as
\begin{align}\label{boxdualnotation}
I_{\rm box} = \int \frac{d^{D}y}{i \pi^{D/2}} \frac{1}{(y-y_{1})^2 (y-y_{2})^2 (y-y_{3})^2 (y-y_{4})^2 }\,,
\end{align}
where now $s=y_{13}^2$ and $t=y_{24}^2$, with $y_{ij}=y_i - y_j$.

This notation is very convenient, as it allows us to give a very simple expression for the Feynman
parametrization of such integrals.
Let us consider a general one-loop integral in the dual notation, 
\begin{align}
I_{n} = \int \frac{d^{D}y}{i \pi^{D/2}} \prod_{i=1}^{n} \frac{1}{[-(y-y_{i})^2 + m_{i}^2]^{a_i}}\,,
\end{align}
where we have allowed the propagators to depend on masses $m_{i}$, and to be raised to arbitrary powers $a_{i}$.

 Introducing Feynman parameters in the usual way, performing a Wick rotation to Euclidean space, and carrying out the $D$-dimensional integration using a Gaussian integral, one arrives at
 \begin{align}\label{generaloneloop}
I_{n} = \frac{\Gamma(a-D/2)}{\prod_i \Gamma(a_{i})} \int_0^\infty  \prod_{i=1}^{n} d\alpha_{i} \alpha_{i}^{a_i -1} \frac{ U^{a-D} \delta(\sum_{i}  \alpha_i -1) }{[ V + U \sum_{i=1}^{n} \alpha_i m_{i}^2 -i0 ]^{a-D/2}}\,,
 \end{align}
with
\begin{align}
V = \sum_{i<j} \alpha_i \alpha_j (-y_{ij}^2) \,, \qquad 
U = \sum_{i} \alpha_i \,, \qquad
a = \sum_{i} a_i \,.
\end{align}
This formula is derived for integer dimension $D$ and exponents $a_{i}$, in a region where the Feynman integral converges. We will use eq. (\ref{generaloneloop}) as the definition for integrals
with other values of these parameters, via analytic continuation.

At $L$ loops, there is a generalization of eq. (\ref{generaloneloop}),
 \begin{align}\label{generalmultiloop}
I_{n,L} = \frac{\Gamma(a- L D/2)}{\prod_i \Gamma(a_{i})} \int_0^\infty  \prod_{i=1}^{n} d\alpha_{i} \alpha_{i}^{a_i -1} \frac{ U^{a-(L+1)D/2} \delta(\sum_{i}  \alpha_i -1) }{[ V + U \sum_{i=1}^{n} \alpha_i m_{i}^2 ]^{a-L D/2}}\,,
 \end{align}
with $U$ and $V$ being homogeneous polynomials in the $\alpha$ parameters
 that have a graph theoretical definition, see e.g. \cite{Smirnov:2004ym}.

{\bf Example: One-loop on-shell box integral in $D=6$ dimensions.}\\
In this simple case, we can directly carry out the Feynman parameter integrations.
Applying the general one-loop formula (\ref{generaloneloop}) with $D=6$, we obtain
\begin{align}
I_{\rm box}^{D=6} = \int \prod_{i=1}^{4} d\alpha_i \, \frac{\delta(\sum \alpha_i -1)}{ \alpha_1 \alpha_3 (-s) + \alpha_2 \alpha_4 (-t) }\,.
\end{align}
Here we assume $s<0,t<0$, which allows us to drop the Feynman $i0$ prescription.
Changing variables according to $\alpha_1 = w x  z$, $\alpha_2 = w (1-x) z$, $\alpha_3 = w y (1-z)$, $\alpha_4 = w (1-y)(1-z)$ with Jacobian $w^3 z (1-z)$, this is easily evaluated, with the result
\begin{align}\label{box6dresult}
I_{\rm box}^{D=6} = \frac{1}{2} \frac{1}{s+t} \left( \log^2 \frac{s}{t} + \pi^2 \right) \,.
\end{align}

\subsection{Infrared and ultraviolet divergences}

Feynman integrals occurring in quantum field theory can have two types of divergences 
having a specific physical origin. The first are ultraviolet (UV) divergences, related to
the region of large loop momentum $k$.
The degree of UV divergence of an integral can be easily determined via power counting.
For example, in the large momentum region the one-loop integral $I_n$ can be approximated
by 
\begin{align}
I_{n} \sim \int \frac{dR}{R} R^{D-2a}\,,
\end{align}
for $R=|k| \gg 1$, so that a divergence occurs if $D<2a$.
Therefore, bubble integrals with $a_{1}=a_{2}=1$ in four dimensions
have a logarithmic divergence, while triangle and box integrals are UV finite, for example.
This explains why the box integral $I_{\rm box}^{D=6}$ considered above is UV finite.
 
Feynman integrals in on-shell kinematics can have another type of singularity,
infrared divergences, that originate from different loop integration regions.
By infrared divergences we mean both soft and collinear divergences. The former
originate from integration regions where $k \ll 1$, while the latter come from regions where
the loop momentum becomes collinear to an on-shell external momentum, $k \sim p$.

Both UV and IR divergences can be regulated using dimensional regularization.

\subsection{Behavior near singular points}
We will be interested in computing Feynman integrals as a function of kinematic invariants 
(e.g. $s$, $t$, or masses) and of the space-time dimension $D=4-2 \epsilon$. In order to know what class of functions to expect it is important to think about the asymptotic behavior of the integrals in (potentially) singular limits.

A typical example is a Regge limit, where $s \gg t$, where one expects in general divergences
of the form $s^{-p} \log^q s/t$. Cf. eq. (\ref{box6dresult}) for an example.

The fact that the functions we are computing have an integral representation such as eq. (\ref{generalmultiloop}) restricts the type of singular behavior that we can obtain.
E.g., in the Regge limit, the leading behavior can be predicted by finding the dominant region in the space of Feynman parameter integration \cite{bookEden}.
It is rather obvious from such an analysis that an integral will be bounded in a limit
by a power with a certain exponent.
This property means that we are dealing with integrals with regular singularities only
(i.e., no essential singularities).
Moreover, the special dependence of  the Feynman parameter integral formula (\ref{generalmultiloop}) 
suggests that the scaling exponents that can appear are linear in the space-time dimension.
In fact, the analysis of regions can be used to determine the different scaling exponents \cite{Beneke:1997zp}.

In summary, we have the following important properties of Feynman integrals,
\begin{itemize}
\item Feynman integrals only have regular singularities in the kinematic variables.
\item The scaling exponents near a singularity are linear in the space-time dimension $D$.
\end{itemize}
In the section \ref{section_algorithm}, we will analyze differential equations satisfied by Feynman
integrals. We will see that these properties imply that a certain 
canonical form of the differential equations can be reached in an algorithmic way.

\section{Integral families and differential equations: an invitation}
\label{section_de}

In this section we explain the following concepts:
Given a Feynman graph (and corresponding Feynman integral), we define a family
of Feynman integrals associated to it. This family consists of, roughly speaking,
all Feynman graphs with the same propagator structure, but arbitrary powers of the propagators.
This includes cases with fewer propagators, i.e. subgraphs.
We derive linear identities between elements of such family, which imply the notion of a basis.
Finally, we explain how to derive differential equations in the external invariants for the basis integrals.

\subsection{Integral families and basis}

\begin{figure}[t]{
\begin{center}
\includegraphics[width=0.35\textwidth]{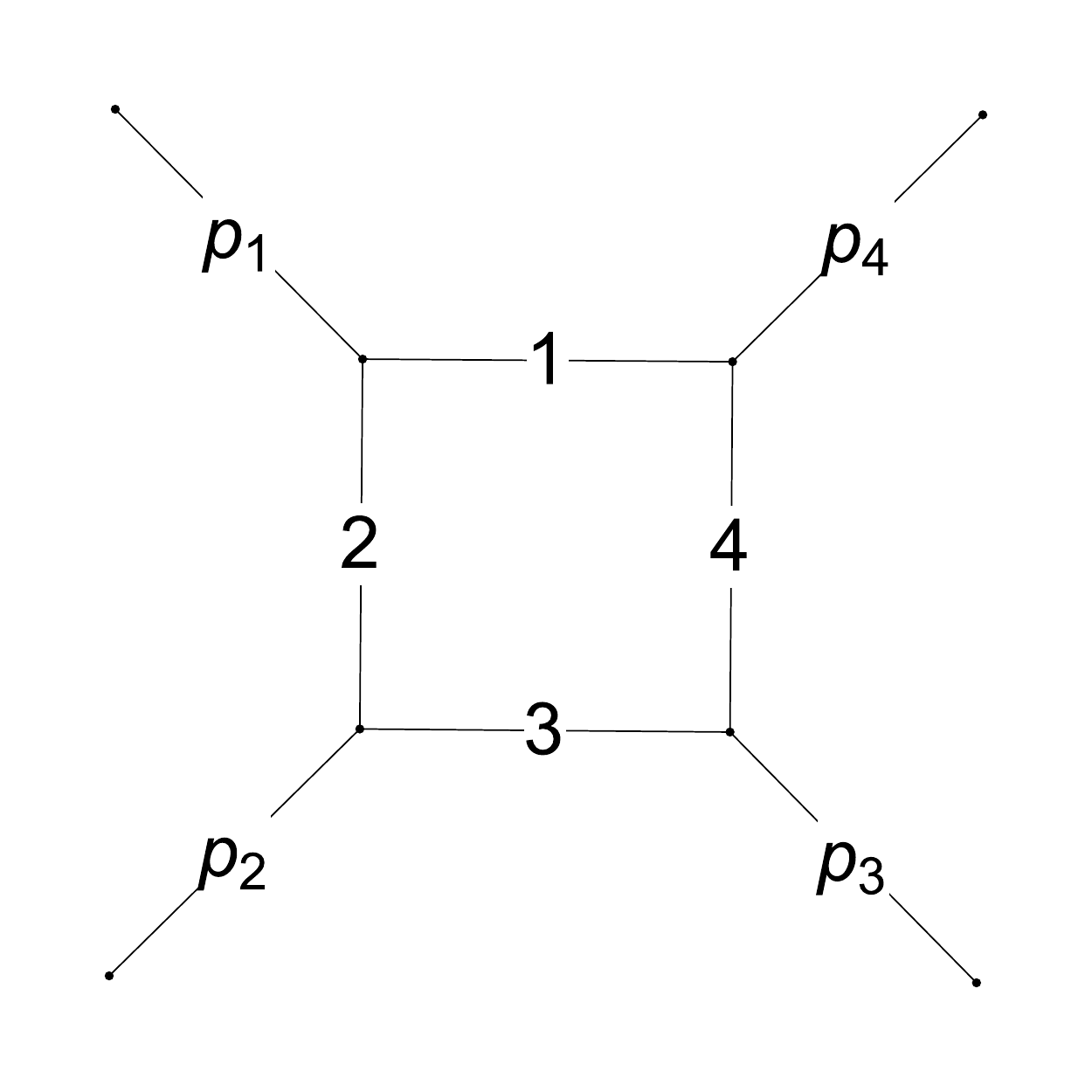}
\end{center}
\caption{One-loop box integral family considered in the main text.}
\label{fig:example1loopbox}
}
\end{figure}

To illustrate the ideas we will proceed with the example of the one-loop box integral
considered in eq. (\ref{boxdualnotation}). 
The first step consists in generalizing it to arbitrary (integer) powers of the propagators,
\begin{align}\label{boxdualnotation2}
G_{a_1,a_2,a_3,a_4}(D;s,t) = \int \frac{d^{D}y}{i \pi^{D/2}} \prod_{i=1}^{4} \frac{1}{[-(y-y_{i})^2 ]^{a_{i}}}\,,
\end{align}
see Fig. \ref{fig:example1loopbox}.
We recall that due to the on-shell conditions, we have $y_{12}^2 =y_{23}^2 = y_{34}^2 =y_{41}^2=0$,
and the integral depends on $s=y_{13}^2$ and $t=y_{24}^2$ (and on the dimension $D$).

For positive $a_{i}$, this is a box integral, with propagators raised to general powers.
If one of the $a_{i}$ is zero, we have a triangle integral, etc. Negative values of the $a_{i}$ correspond
to numerator factors. We call the set of $G$ for arbitrary integer powers of the $a_{i}$ an integral
family (associated to the box diagram).
We will see presently that this notion is useful to understand the structure of the differential
equations satisfied by this integral.

The integrals in a given family are in general not independent. There are linear relations that have a very simple
origin, namely integration by parts (IBP) relations \cite{Chetyrkin:1981qh}.
These identities follow from the fact that total derivatives vanish in dimensional regularization.\footnote{Of course it is also possible to work with IBP identities that include boundary terms.}
Therefore we have e.g.
\begin{align}\label{ibp1}
0 = \int \frac{d^{D}y}{i \pi^{D/2}} \frac{\partial}{\partial y^{\mu}} \xi^{\mu} \prod_{i=1}^{4} \frac{1}{[-(y-y_{i})^2 ]^{a_{i}}}\,,
\end{align}
where $\xi$ is some vector, e.g. $y-y_{1}$.

Acting with the differential operator on the rest of the integrand, 
and performing a little amount of algebra,
one observes that the r.h.s. of eq. (\ref{ibp1}) can be expressed in terms of members of the integral family,
albeit with different values of $\{a_{1}, \ldots a_{4}\}$. 
The coefficients in this relation are rational functions of $s,t$, and $D$.
Introducing the operators $Y_{i}^{\pm}$, with $Y_{1}^{\pm} G_{a_1,a_2,a_3,a_4} = G_{a_1 \pm 1,a_2,a_3,a_4}$, etc., we have the IBP relations
\begin{align}\label{ibp2}
0 =\left[ (D-2 a_1 -a_2 -a_3-a_4) - s a_3 Y_3^+  + (-a_2 Y_2^+  -a_3 Y_3^+ -a_4 Y_4^+) Y_1^-  \right]  G_{a_1,a_2,a_3,a_4}\,,
\end{align}
as well as similar relations obtained by cyclic symmetry.
We note that the relations are linear in the integrals $G$, as promised.

The IBP identities relate integrals having different indices $a_{i}$. For integer
values of the $a_i$, one can devise a simple strategy to reduce any integral to a 
known basis \cite{Kotikov:1990kg}. This is done by noticing that the relations (\ref{ibp2}) can be used
to reduce the value of $a = \sum_{i=1}^{4} a_{i}$. This is done until one of the indices is zero,
and then repeated for the remaining indices.
In this way one sees that one can relate any integral in the family to a basis.
In the present case, this basis consists of three elements, which can be chosen
to be the $s$- and $t$- channel bubble integrals, and the box integral, with unit powers:
$G_{0,1,0,1}, G_{1,0,1,0}, G_{1,1,1,1}$.

Let us give two examples of the integral reduction.
\begin{align}
G_{2,1,1,1} =& \frac{D-5}{s} G_{1,1,1,1} - \frac{4 (D-5)(D-3)}{(D-6)s t^2} G_{0,1,0,1} \,, \\
G_{1,1,0,1} =& \frac{2 (D-3)}{(D-4) t} G_{0,1,0,1} \,.
\end{align}
The first relation demonstrates how higher powers in the denominators can be removed, as explained above. The second example shows that the on-shell triangle is trivially related to the bubble integral,
which explains the absence of triangle integrals in the above basis.

In general, the IBP relations imply the existence of a finite basis of integrals for the family under consideration.
In the literature the basis integrals are often referred to as master integrals. We prefer the
word basis, as it reminds one that in linear algebra, there is a freedom in the choice of basis.
This will be very important in the following sections.

Here we considered a simple example, but the properties observed turn out to be completely 
general. Let us summarize the main points.
When studying a Feynman integral, it is useful to consider the family of Feynman 
integrals associated to it. Integrals in this family satisfy relations following from integration-by-parts identities. 
The latter are linear in the integrals, and rational in the kinematic variables and the dimension $D$.
They imply that the family has a finite basis.

At one loop it is possible to solve the IBP relations directly and to write down a relation between arbitrary
integrals and the chosen set of basis integrals. At higher loops, such a general solution is not known, so that
for each case under consideration one usually constructs a  sufficient number of IBP relations.
Although straightforward in principle, this is rather tedious to do by hand, so that one usually 
employs some appropriate computer algebra program. Various implementations exist, see \cite{Anastasiou:2004vj,Smirnov:2014hma,vonManteuffel:2012np,LiteRed}.

\subsection{Differential equations}

One result of the previous section was that for a given family of Feynman integrals there exists
a finite-dimensional basis $\vec{f}$. In practice, the latter is found by writing down a sufficient number of IBP relations.
Given such a basis, any integral in the family can be written in terms of a linear combination of basis 
integrals, with rational prefactors in the kinematic variables and $D$. Therefore it is sufficient
to compute the basis integrals. We will use differential equations in the kinematic variables to that end.

In the example of the one-loop box integral family, the kinematical variables are $s$ and $t$.
We implement differential operators $\partial_s$ and $\partial_t$ acting on the integral representation (\ref{boxdualnotation2}), which depends on the vectors $y_i$ (or, equivalently, on the $p_i$), via the chain rule.
When doing so one has to make sure that the differential operators commute with the on-shell and momentum conservation constraints.
 
For an analogy, think of a two-dimensional space with constraint $x^2+y^2=1$, i.e. points lying on the unit circle. In that case, only the operator $y \partial_x - x \partial_y$ is allowed. Of course, one could also introduce radial coordinates $x = r \cos \theta, y = r \sin \theta$, so that the constraint becomes $r^2=1$, which means that one can freely vary $\theta$.
For on-shell massless scattering amplitudes, likewise, one can solve the momentum conservation and on-shell constraints using momentum twistor variables \cite{Hodges:2009hk}. 
Here we will use the momentum-space variables and construct differential operators commuting with the constraints.

{\bf Example: differential operators}\\
Let us use the momentum-space notation for the family of one-loop box integrals, and eliminate $p_{4}$ using momentum conservation,
\begin{align}
G_{a_1,a_2,a_3,a_4} =  \int \frac{d^{D}k}{i \pi^{D/2}} \frac{1}{[-k^2]^{a_1} [-(k+p_{1})^2]^{a_2}[- (k+p_1 +p_2)^2]^{a_3}[- (k+p_1+p_2+p_3)^2 ]^{a_4}}\,,
\end{align}
Let us construct a differential operator for $\partial_s$ that can act on the r.h.s. of this equation.
This is easily achieved by making the ansatz
\begin{align}\label{ansatzdiffs}
\partial_s = (\beta_1 p_1 + \beta_2 p_2 + \beta_3 p_3)\cdot \partial_{p_{1}}  \,.
\end{align} 
Imposing that this operator should commute with the on-shell conditions $p_1^2=0$ and $(p_1 + p_2 + p_3)^2=0$, and imposing the normalization condition $\partial_s (p_1 + p_2)^2 =1$ fixes the parameters in eq. (\ref{ansatzdiffs}) 
to be
\begin{align}
\beta_1 = \frac{2 s+t}{2 s (s+t)} \,,\qquad \beta_2 = \frac{1}{2 s} \,,\qquad \beta_3 = \frac{1}{2 (s+t)} \,.
\end{align}

When acting with such differential operators on the Feynman integral representation,
we have to perform algebraic manipulations similar to those when deriving the IBP relations.
It is clear that one obtains integrals within the same family of integrals.
The fact that there is a basis means that we can rewrite the result of the differentiation as a linear combination
of basis integrals.
In other words, we have
\begin{align}\label{general_DE_st}
\partial_s \vec{f}(s,t;\epsilon) = A_s(s,t,\epsilon) \vec{f}(s,t;\epsilon) \,,\\
\partial_t \vec{f}(s,t;\epsilon) = A_t(s,t,\epsilon) \vec{f}(s,t;\epsilon) \,.
\end{align} 
where $A_{s}$ and $A_{t}$ are $N$ by $N$ matrices, with $N$ being the number of 
basis integrals $\vec{f}$.
By construction, they contain only rational functions of $s,t,\epsilon$ as entries.

In other words, Feynman integrals satisfy first-order systems of (partial) differential equations.
The matrices $A_{i}$ can be computed algorithmically, as outlined in this section.

{\bf Example: Differential equations for the family of one-loop $2 \to 2$ integrals.}\\
We already saw that in this example there are three basis integrals.
Integral reduction suggests the following basis choice,
\begin{align}\label{basischoicebox0}
f_1 =&  G_{0,1,0,1}\,, \nonumber \\
f_2 =&  G_{1,0,1,0} \,, \\
f_3 =&  G_{1,1,1,1} \,. \nonumber
 \end{align}
 With this choice, we find the following matrices in eq. (\ref{general_DE_st}),
 \begin{align}
 A_s = \left( \begin{array}{ccc}
0 & 0 & 0 \\
0 & -\frac{\eps}{s} & 0 \\
\frac{-2 (1-2 \eps)}{s t (s+t)} & \frac{2(1-2 \eps)}{s^2(s+t)} & -\frac{s+t + \eps t}{s (s+t)} \end{array} \right) \,,\qquad 
A_t  =
 \left( \begin{array}{ccc}
-\frac{\eps}{t} & 0 & 0 \\
0 & 0 & 0 \\
\frac{-2 (1-2 \eps)}{t^2 (s+t)} & \frac{-2(1-2 \eps)}{s t (s+t)} & -\frac{s+ \eps s+  t}{t (s+t)} \end{array} \right) \,.
\end{align}
We can make the following observations.
\begin{itemize}
\item Computing $s A_s + t A_t  = {\rm diag}(-\eps, -\eps ,-2-\eps)$, the scaling dimensions of the
integrals are correctly reproduced. We can set them to zero by choosing appropriate dimensional normalization factors, so that we only have one non-trivial variables $x=t/s$.
\item The equations for the bubble integrals $f_1$ and $f_2$ are trivial, and indeed being single-scale integrals, their functional dependence follows from dimensional analysis.
\item The equations have the singular points $s=0$, $t=0$, $s = \infty$, $t=\infty$, and $s=-t$ (i.e. $u=0$). The latter singularity may be surprising for planar integrals, and as we will see occurs only after analytic continuation.
\end{itemize}

As a preview of the general method to be discussed in the following sections, 
let us make the following educated basis choice (to be justified later),
\begin{align}
g_1 =& c (-s)^\eps  t G_{0,1,0,2}\,, \nonumber \\
g_2 =& c (-s)^\eps  s G_{1,0,2,0} \,, \label{eqbasisbox} \\
g_3 =& c \eps (-s)^\eps s t G_{1,1,1,1} \,, \nonumber 
 \end{align}
 with $c=\eps e^{\eps \gamma_{\rm E}}$ being a normalization factor, and with $\gamma_{E}$ being Euler's constant.
The $g_i$ are chosen to be dimensionless, such that they depend on $x$ and $\eps$ only.
Implementing the derivative $\partial_s$ as explained above, and using the chain rule, we find
\begin{align}\label{DEoneloopbox}
\partial_x \vec{g}(x;\eps) = \eps \left[ \frac{a}{x} + \frac{b}{1+x} \right] \vec{g}(x,\eps) \,,
\end{align}
where
\begin{align}
a = \left( \begin{array}{ccc}
-1 & 0 & 0 \\
0 & 0 & 0 \\
-2 & 0 & -1 \end{array} \right) \,,\qquad
b = \left( \begin{array}{ccc}
0 & 0 & 0 \\
0 & 0 & 0 \\
2 & 2 & 1 \end{array} \right)  \,.
\end{align}
The system (\ref{DEoneloopbox}) can be solved easily in an expansion in $\eps$.
One sets 
\begin{align}
\vec{g} = \sum_{k \ge 0} \eps^k \vec{g}^{(k)}(x) \,,
\end{align}
and plugging this into eq. (\ref{DEoneloopbox}) it becomes clear that at each order
in $\eps$, the r.h.s. of that equation is known and can be integrated.

Let us discuss the boundary conditions for the equations.
As already discussed, the bubble integrals are trivially known:
a short calculation using the formulas of section \ref{section_intro} shows that they are given
by
\begin{align}\label{formula_bubble}
G_{a_1,0,a_2,0} = (-s)^{D/2 -a} \frac{\Gamma(a-D/2) \Gamma(D/2-a_1) \Gamma(D/2-a_2)}{\Gamma(a_1) \Gamma(a_2) \Gamma(D-a)} \,,
\end{align}
with $a=a_1 + a_2$.
In application to our case, we have
\begin{align}
g_1 =   x^\eps g_2 \,, \qquad 
g_2 = - e^{\eps \gamma_{\rm E}}  \frac{\Gamma^2(1-\eps) \Gamma(1+\eps)}{\Gamma(1-2 \eps)}     \,.
\end{align}
Finally, we need a boundary condition for $g_3$. We can use the fact that planar integrals should not have $u$-channel singularities, which implies that $g_3$ should stay finite as $x\to -1$, despite the presence of the matrix $b$ in eq. (\ref{DEoneloopbox}).

This fixes the solution to all orders in the $\eps$ expansion. 
The first few orders are given by
\begin{align}
g_3 =&  4  + \eps \left[ -2 \log x \right] +\eps^2 \left[-\frac{4 \pi^2}{3} \right] 
+\eps^3 \left[ \frac{7 \pi^2}{6} \log x + \frac{1}{3} \log^3 x - \pi^2 \log(1+x)   \right.
\nonumber \\ & \left. - \log^2 x\log(1+x) -2 \log x {\rm Li}_{2}(-x) +2 {\rm Li}_{3}(-x) - \frac{34}{3} \zeta_3 \right] + \cO(\eps^4)\,,
\end{align}
where ${\rm Li}_{n}$ is a polylogarithm, defined by
\begin{align}
{\rm Li}_1 (x) = -\log(1-x) \,,\qquad x \, \partial_x {\rm Li}_n(x) = {\rm Li}_{n-1}(x) \,,\quad n>1 \,,
\end{align}
and ${\rm Li}_{n}(0)=0$.
In section \ref{section_canonical} we will discuss a more general class of functions that is useful
for writing the solutions to such differential equations.

We note that one can associate a notion of weight to the above functions, 
corresponding to the number of integrations. For example ${\rm Li}_{n}$ has weight $n$.
Moreover, if one also associates weight $-1$ to $\eps$, then one observes that the $g_i$ have 
uniform weight $0$ for all terms in the $\eps$ expansion.

The main conclusion of this section is that the basis integrals satisfy a linear system of differential
equations with rational coefficients. This follows simply from the existence of a basis, and the
way the derivative operators act on integrals of a given family.

\subsection{Singular points of the differential equations}

In section \ref{section_intro} we observed general properties of Feynman integrals near their singular points.
Let us study the one-loop box example from this point of view.
We see that eq. (\ref{DEoneloopbox}) has the singular points $x=-1,0,\infty$.
(The analysis of the point at infinity can be treated as the other ones after 
performing an inversion.)
What is their physical meaning?
It is easy to see that they correspond to the limits $u = -s-t \to0, t\to 0$, and $s\to 0$, respectively.

The differential equations tell us how the solutions behave near those points.
Let us consider the limit $x\to 0$. Keeping only the leading term on the r.h.s., 
we find the solution
\begin{align}
\lim_{x \to 0} \vec{f}(x,\eps) = x^{\eps a} \vec{f}_{0}(\eps) \,, 
\end{align}
where $f_{0}(\eps)$ is a boundary vector.
The matrix exponential evaluates to
\begin{align}\label{examplematrixexp}
x^{\eps a} = \left( \begin{array}{ccc}
x^{-\eps} & 0 & 0 \\
0 & 1 & 0 \\
-2 \eps x^{-\eps} \log x & 0 & x^{-\eps} \end{array} \right)\,.
\end{align}
This illustrates the statements at the end of section \ref{section_intro}.
The solutions are linear combinations of different terms $x^{\alpha}$, 
where $\alpha$ are linear in the dimension. 
We note that the $\alpha$ are the eigenvalues of the matrix multiplying the singular point,
in this case $\eps a$.

This example illustrates that in general there are two sources
of logarithms in the expansion around a singular point. The first is the matrix
exponential itself, c.f. eq. (\ref{examplematrixexp}), and the second
is the expansion for small $\eps$ of exponentials such as $x^{-\eps}$.

In general, the structure of singularities will not always be as manifest
as in this example.
To understand this better, consider the scalar differential equation
\begin{align}
\partial_x f(x) = a/x^2 f(x)\,,
\end{align}
which is more singular at $x=0$ compared to the previous case.
Indeed, the solution $f(x) = e^{-a/x} f_{0}$ has an essential singularity at $x=0$.
Such a function cannot appear in individual Feynman integrals.

Can one conclude therefore that the differential equations contain simple poles only?
Unfortunately, this conclusion would be premature, since the matrix nature of the equations allows for `spurious'
terms to occur.
This fact can be seen from the following example,
\begin{align}
\partial_x \vec{f}(x,\eps) =  
\left( \begin{array}{cc}
\frac{\eps}{x}  & 0 \\
-\frac{1}{x^2} & \frac{\eps}{1+x}  
\end{array} \right)  \vec{f}(x,\eps)\,.
\end{align}
Here the $1/x^2$ term is spurious. It can be removed
by a simple change of basis,
\begin{align}
\vec{f} = T \vec{g}  \,,\qquad T =  \left( \begin{array}{cc}
1  & 0 \\
\frac{1}{(1-\eps) x} & \frac{1}{1-\eps}  
\end{array} \right) \,,
\end{align}
which leads to
\begin{align}
\partial_x \vec{g} =  
\eps
 \left[ \frac{1}{x} \left( \begin{array}{cc}
1  & 0 \\
1 & 0 
\end{array} \right) 
+ \frac{1}{1+x} \left( \begin{array}{cc}
0  & 0 \\
-1 & 1 
\end{array} \right) \right] 
\vec{g}\,.
\end{align}

The reason the singularity structure was manifest in the example of the one-loop box integral
crucially had to do with the basis choice made above, namely eq. (\ref{eqbasisbox}) vs. eq. (\ref{basischoicebox0}). Obviously, the two basis choices are related by a transformation $T$.
The next sections are devoted to a better understanding of this point.
In particular, section \ref{section_algorithm} discusses this from the point of view of the DE, 
while section \ref{section_leading_singularities} relates it to properties of the loop integrands.

\section{Algebraic simplifications of differential equations}
\label{section_algorithm}

Here we discuss how the expected singularity structure
can be made manifest, leading to a canonical form of 
the differential equations.
In the physics literature, some of these ideas were
presented e.g. in \cite{Henn:2013pwa,Argeri:2014qva,Caron-Huot:2014lda,Gehrmann:2014bfa,Hoschele:2014qsa}.
We also discuss how iterated integrals, and
more complicated functions such as elliptic functions occur.

We saw in the previous section that integrals for families of Feynman integrals satisfy first-order coupled systems
of differential equations. Here we will focus on the case where the integrals 
depend on one kinematical variable, $x$, for simplicity.
This is the case of the example discussed above, since one can always normalize the integrals 
to remove one of the scales.
Then, for some choice of basis $\vec{f}$ we have
\begin{align}\label{DEgeneric}
\partial_{x} \vec{f}(x;\epsilon) = A(x,\epsilon) \, \vec{f}(x;\epsilon)\,,
\end{align}
where $A$ is an $N\times N$ matrix.
{}From the structure of the IBP relations it follows that $A$ depends on $x$ and $\epsilon$ in a rational way\footnote{For the time being, we choose a basis $\vec{f}$ without introducing non-rational dependence on $x$.}.

\subsection{Simplifying the dependence on $x$}

The singularities of the differential equations (\ref{DEgeneric}) have to correspond to singularities of the original Feynman integrals. If we denote the singular points by $x_{k}$, we expect the leading behavior of the integrals to be given by terms that grow like $\sim (x-x_{k})^{\alpha}$, for some values of $\alpha$. This is to say that (\ref{DEgeneric}) should only have regular singularities, i.e. it should be a Fuchsian system of differential equations.

To specify what this means in terms of the system of DE  (\ref{DEgeneric}), we first have to recall
that there is a gauge degree of freedom in the choice of the basis.
Indeed, starting from eq. (\ref{DEgeneric}), we can switch basis according to
\begin{align}
\vec{f} = T \vec{g},
\end{align}
for some invertible matrix $T$, which transforms (\ref{DEgeneric}) into an equivalent
system,
\begin{align}
\partial_{x} \vec{g}(x;\epsilon) = B(x,\epsilon) \, \vec{g}(x;\epsilon)\,.
\end{align} 
where 
\begin{align}
B = T^{-1} A T - T^{-1} \partial_x T\,.
\end{align}
Let us now investigate the DE near a singular point. We take $x=0$ without loss of generality.
The matrix $A$ in eq. (\ref{DEgeneric}) has the expansion\footnote{For simplicity of notation, we will continue using the letter $A$, even if it was obtained by a gauge transformation from the original system.}
\begin{align}\label{expandA}
A(x,\epsilon) = \frac{1}{x^p} \sum_{k \ge 0} x^k A_{k}(\eps) \,,
\end{align}
for some value of $p$.
The system of DE is regular singular in $x=0$ if there exists some gauge transformation $T$,
for which the matrix $A$ appearing in the DE has $p\le1$ (if $p <1$ the solution has no singularity at $x=0$).

In other words, the fact that (\ref{DEgeneric}) is a Fuchsian system implies that for each singular point, one can find a gauge transformation $T$ such that the gauge equivalent system has a matrix which has the leading behavior
\begin{align}
A(x,\eps) =  \frac{1}{x} A_0(\eps)  + \cO(x^0)\,,
\end{align}
i.e. with $p\ge 1$ in eq. (\ref{expandA}).
As a consequence, near each singular point the solution behaves as $x^{A_{0}(\eps)}$.

The degree of singularity of a system of differential equations was studied by Moser \cite{Moser,wasow1965asymptotic}.
It was shown that under certain conditions on the two leading matrices in the expansion
(\ref{expandA}) near a singular point, the order of the singular term can be reduced. 
It is important to note that the necessary transformation is rational in $x$, see e.g. \cite{Barkatou95,pfluegel}. 
This implies that removing spurious singularities at one singular point does not influence the behavior at other points (except possibly at infinity).
Therefore one can algorithmically construct a rational matrix $T$ such that the DE system reads
\begin{align}\label{canonical1}
\partial_x \vec{f}(x,\eps) = \left[ \sum_k \frac{a_{k}(\eps)}{x-x_k} + p(x,\eps) \right] \vec{f}(x,\eps) \,,
\end{align}
where $p(x,\eps)$ is polynomial in $x$. If $p(x,\eps) \neq 0$, this means there is still an undesired spurious singularity at infinity.

The question whether $p(x,\eps)$ can be removed without introducing further singular points is related to the Riemann-Hilbert problem, see e.g. \cite{RiemannHilbert}. It is an interesting question how to decide whether this is possible and to construct such a transformation matrix in the positive case.
A more pragmatic solution consists in introducing another singular point, not yet present in the list 
$\{ x_{k} \}$, to ``balance'' the transformation at infinity. In this way, we can use the above algorithm to obtain the form (with different matrices compared to eq. (\ref{canonical1}))
\begin{align}\label{canonical2}
\partial_x \vec{f}(x,\eps) = \left[ \sum_k \frac{a_{k}(\eps)}{x-x_k}  \right] \vec{f}(x,\eps) \,,
\end{align}
This is a system with manifestly only regular singularities.

\subsection{Simplifying the dependence on $\eps$}

We have just explained how for Feynman integrals one can put the differential
equations in the form of eq. (\ref{canonical2}), using the algorithms of refs. \cite{Moser,wasow1965asymptotic,Barkatou95}. 
(We restricted ourselves to the single variable case for the sake of the presentation,
but similar results can be obtained for the multi-variables case.)
This equation makes the behavior of $\vec{f}$ near the singular points in $x$ manifest.
Indeed, the solution takes the form (we again take $x=0$ without loss of generality),
\begin{align}
\vec{f } = P(x,\eps) x^{a_{0}(\eps)} \vec{f}_{0}(\eps) \,,
\end{align}
where $\vec{f}_{0}(\eps)$ is a boundary vector, independent of $x$, at $x \to 0$,
and 
\begin{align}
P(x,\eps) = \mathbb{I} + \sum_{m\ge 1} x^m P_{m}(\eps) \,,
\end{align}
is a matrix polynomial in $x$, whose expansion coefficients $P_{m}(\eps)$ can be determined recursively  from the information in eq. (\ref{canonical2}) \cite{wasow1965asymptotic}.

In other words, this form already contains all the information about the scaling
behavior of the integrals near the singular points. 
In particular, as discussed in section \ref{section_intro}, the 
eigenvalues of $a_{k}(\eps)$ are related to different regions
in integration space, and one expects them to be linear in $\eps$. 

This raises the question whether the dependence on $\eps$ in eq. (\ref{canonical2})  can be simplified.
By construction, we know that $\eps$ only appears in a rational way. Moreover, it is clear that poles in $\eps$ in $a_k$ must be spurious, and these can be removed similarly to the removal of spurious divergences in $x$, see e.g. \cite{wasow1965asymptotic,2014arXiv1401.5438B}.

For a polynomial dependence on $\eps$, we can distinguish between two cases:
if the r.h.s. of eq. (\ref{canonical1}) is $\cO(\eps)$, the solution at each order in $\eps$ 
can be obtained in terms of iterated integrals. If the r.h.s. starts at order $\eps^0$, the solution may be more complicated.

So we see that a crucial question is whether we can construct a transformation
that removes the $\eps^0$ part of the matrix on the r.h.s. of eq. (\ref{canonical2}), and
what the nature of this transformation is.
This is best explained via a few examples.

\begin{itemize}
\item {\bf Case 1:} integrating out the $\eps^0$ term amounts to choosing a rational normalization factor.

As an example, imagine choosing a normalization factor $s^2$ instead of $s \, t$ for $g_{3}$ in eq. (\ref{eqbasisbox}). In that case, one obtains a $\eps^0$ in the DE. The latter can be removed (by construction) by a simple rational transformation $T$.

\item {\bf Case 2:} integrating out the $\eps^0$ term can be done using algebraic functions; sometimes a change of variables leads to a rational dependence.

This is something that occurs typically for integrals involving masses. As an example let us choose the
$2\times 2$ system for a massive bubble and tadpole integral \cite{Henn:2013woa,Henn:2014yza}. The bubble integral depends on an external invariant $s$, as well as on an internal mass $m$. We set $s=x$ and $m^2=1$ without loss of generality. Before choosing appropriate normalization factors, the differential equation in $x$ reads
\begin{align}
\partial_x \vec{f}(x,\eps) = 
 \left( \begin{array}{cc}
0  & 0 \\
\frac{\eps}{4-x} & \frac{2+\eps x}{(4-x) x}  
\end{array} \right) \vec{f}(x,\eps)\,.
\end{align}
Here, integrating out the constant term in $\eps$ amounts to choosing a transformation matrix
$T = {\rm diag}(1,1/\sqrt{1-1/x})$.
Note that contrary to all transformations discussed so far, this transformation is not rational (in the chosen variables).
Under $\vec{f} \longrightarrow T \vec{f}$, the system of DE becomes
\begin{align}
\partial_x \vec{f}(x,\eps) = 
\eps  \left( \begin{array}{cc}
0  & 0 \\
-\frac{1}{\sqrt{x (x-4)}} & \frac{1}{x-4}  
\end{array} \right) \vec{f}(x,\eps)\,.
\end{align}
The r.h.s. is $\propto \eps$, as promised.
Finally, we note that in this case, one can recover a rational form of the equations
by employing the change of variables $x=-(1-y)^2/y$, with the resulting system having
regular singularities in $y=\pm 1, 0,\infty$.

\item {\bf Case 3:} integrating out the $\eps^0$ term leads to elliptic or more complicated functions.

A simple example of this kind is the system of DE satisfied by complete elliptic integrals.
The difference w.r.t. the previous example lies in the fact that this is a coupled $2 \times 2$ system of differential equations, even at $\eps=0$. The fact that the appearance of elliptic functions can be
seen in this way was also pointed out in the algorithm of ref. \cite{Caron-Huot:2014lda}. 

For Feynman integrals, the simplest example where this can occur is perhaps the two-dimensional  two-loop sunrise integral with equal masses, see e.g. \cite{Laporta:2004rb,Remiddi:2013joa,Bloch:2013tra,Adams:2014vja,Bloch:2014qca}.
There are singular points at $x=0,-1,-1/9,\infty$, where $x=m^2/(-p^2)$.
Making an appropriate basis choice, one obtains a system of DE (in $D=2-2\eps$ dimensions) $\partial_x \vec{f}(x,\eps) = A(x,\eps) \vec{f}(x,\eps)$, with the matrix
\begin{align}
A(x,\eps) =& \frac{1}{x}   \left( \begin{array}{ccc}
-2 \eps  & 0 & 0 \\
0 & 1 &  -\frac{1}{4}-\frac{\eps}{2} \\
0  & 0 & 1-\eps 
\end{array} \right)  
+ \frac{1}{1+x}
 \left( \begin{array}{ccc}
0   & 0 & 0 \\
0 & 0 & 0\\
3  & 3+9\eps & -1-2 \eps 
\end{array} \right)  \nonumber \\
& 
+ \frac{1}{1/9+x}
 \left( \begin{array}{ccc}
0   & 0 & 0 \\
0 & 0 & 0\\
-3  & 1+3\eps & -1-2 \eps 
\end{array} \right) \,.
\end{align}
One may verify that all eigenvalues are linear in $\eps$, as expected.
Analyzing this system at $\eps = 0$, and writing it as a second-order eq. for one of the integrals, one recovers the Picard-Fuchs equation discussed e.g. in refs. \cite{Bloch:2013tra,Adams:2014vja}. In other words, integrating out the $\eps^0$ piece of the equation leads to elliptic functions.
This could obviously be generalized to cases where more than two equations are coupled
at $\eps=0$, leading to higher-order Picard-Fuchs equations for the individual integrals.

\end{itemize}

Note added: After these lectures were given, ref. \cite{Lee:2014ioa} appeared, which has some overlap with the material presented in this section. This reference proposes a refined version of balancing the transformations of the Moser algorithm, and, for the case of a Fuchsian system with eigenvalues already normalized to $\mathcal{O}(\eps)$, it gives a formula for constructing a transformation matrix to simplify the $\eps$ dependence of the differential equations. 
The methods presented here and in ref. \cite{Lee:2014ioa} allow for a systematic, albeit not fully algorithmic, simplification of differential equations. 
Challenges for an algorithmic implementation include reaching a Fuchsian form without spurious singularities, i.e. the step from eq. (\ref{canonical1}) to eq. (\ref{canonical2}),
and dealing with non-rational dependence on variables. 
As we have seen above, the latter can appear when integrating out $\eps^0$ terms, 
or, equivalently, when normalizing eigenvalues \cite{Lee:2014ioa}. 
Another open problem is the extension to multi-variable cases. 
In section \ref{section_leading_singularities}, we present an alternative method for constructing an integral basis. This method, based on insights from the singularity structure of Feynman integrands, is very simple to use and has already been successfully applied to complicated multi-variable cases with algebraic dependence on the variables. Also, it typically gives much simpler expressions for the integral basis. Of course, the ideas of this section and section \ref{section_leading_singularities} are complementary, and having both at one's disposal allows one to solve very complicated problems.

\section{Iterated integrals from differential equations}
\label{section_canonical}

In the previous section we discussed how differential equations can be simplified systematically
both in $x$ and in $\eps$ in an algebraic way. Before discussing a completely orthogonal
approach to this, that makes direct use of properties of the Feynman loop integrand, we wish to discuss properties of the solutions of the differential equations.

\subsection{Canonical form of differential equations}

Here we will discuss the case where the differential equations can be put into the form suggested in
ref. \cite{Henn:2013pwa}.
Let us start with the case of one non-trivial variables $x$. The form of the equations proposed in ref. \cite{Henn:2013pwa} reads 
\begin{align}\label{canonicalform1}
d\, \vec{f}(x,\eps) = \eps \, \left( d \, \tilde{A} \right) \,\vec{f}(x;\eps) \,,
\end{align}
with
\begin{align}\label{canonicalform2}
\tilde{A} = \left[ \sum_{k} A_{k} \log \alpha_{k}(x) \right]\,.
\end{align}
In the case where this representation can be reached using only rational transformations (for some choice of variables, see the examples at the end of the previous section), the $\alpha_k$ are
rational functions, i.e.
\begin{align}
\alpha_{k} = x-x_{k} \,,
\end{align}
where the $x_k$ are the locations of the singularities.
In the general case they can depend algebraically on $x$.

The generalization to the multi-variable case is straightforward,
we can simply replace $x$ by $\vec{x}$ in eqs. (\ref{canonicalform1}) and (\ref{canonicalform2}).

The canonical form of eqs. (\ref{canonicalform1}) and (\ref{canonicalform2}) 
has the virtue that the information about the functions $\vec{f}$ is encoded in a 
minimal way: the alphabet $\alpha$ specifies which class of generalized functions
will be required for writing down the answer. As we will see presently, 
the coefficient matrices for each letter determine which linear combination of those
functions is required.  Indeed, the general solution to eqs. (\ref{canonicalform1}) and (\ref{canonicalform2}) in the multi-variable case can be written as Chen iterated integrals \cite{Chen}
\begin{align}\label{general_solution}
\vec{f}(\vec{x},\eps) = {\mathbb{P}} \exp \left[ \eps \int_{ \gamma}  d \, \tilde{A} \right] \vec{f}_{0}(\eps) \,, 
\end{align}
where $ {\mathbb{P}}$ stands for path ordering along the integration contour $\gamma$,
and $\vec{f}_{0}(\eps)$ is a boundary value. This formula is to be understood in an expansion in $\eps$, where the $k$-th term in the expansion is a $k$-fold iterated integral (along $\gamma$).

Let us be more specific about the notation, following closely the recent lecture notes \cite{Brown1,Brown2,Caron-Huot:2014lda} on iterated integrals. We denote by $\mathcal{M}$ the space of kinematical variables, 
and let $\omega_i$ be some differential one-forms (corresponding to entries of $d\, \tilde{A}$).
Moreover, define the pull-back of the differential forms to the unit interval $[0,1]$ via
\begin{align}
\gamma^{*}(\omega_{i}) = k_{i}(t) dt\,.
\end{align}
Then, an ordinary line integral is given by
\begin{align}
\int_{\gamma} \omega_1 = \int_{[0,1]}\gamma^{\star}(\omega_1) = \int_0^1 k_{1}(t_1) dt_1 \,.
\end{align}
The iterated integral of $\omega_1, \ldots \omega_n$ along $\gamma$ is defined by
\begin{align}\label{iterateddef3}
\int_{\gamma} \omega_1 \ldots \omega_n = \int_{0\le t_1 \le \ldots \le t_n \le 1} k_{1}(t_1) dt_1 \ldots k_{n}(t_n) dt_{n} \,.
\end{align}
Iterated integrals have many nice properties, see \cite{Brown1,Brown2}. 
Moreover, the iterated integrals appearing in eq. (\ref{general_solution}) are homotopy invariant (on $\mathcal{M}$ with the set of singularities removed).
This property makes them very flexible, and we will see later that how to rewrite them in terms of more familiar functions, if desired.
The freedom in choosing the integration 
contour $\gamma$ is also useful e.g. for analytic continuation, or for writing integral
representations that have certain desired properties. 

It is important to emphasize the simplicity of the solution (\ref{general_solution}).
Recall the notion of weight introduced earlier, corresponding to the number 
of iterated integrations. 
It is obvious from eq. (\ref{general_solution}) that each term in the $\eps$-expansion is a $\mathbb{Q}$-linear combination of iterated integrals of the same weight. This is a remarkable property that is not true
in a generic integral basis, where results would look far more complicated, with
terms of different weights being mixed, and where prefactors are in general 
algebraic functions of the kinematic variables.
Finally, note that if $\eps$ is assigned weight $-1$ (and defining the weight of a
product as the sum of individual weights), then
one can say that (\ref{general_solution}) has uniform weight zero.

These remarkable simple properties will in fact be a guiding principle for finding an 
appropriate integral basis in section \ref{section_leading_singularities}.
As we will see there, it is possible to anticipate these properties by inspection of the
Feynman integrand, i.e. even before carrying out any integrations.

\subsection{Fixing the boundary conditions}

The system of first-order differential equations (\ref{canonicalform1}), (\ref{canonicalform2}) determines the answer up to an integration constant. In principle, the latter can be fixed by an independent (and simpler) calculation at a preferred kinematic point. However, in practice
the differential equations themselves, together with insight coming from the original Feynman integral representation, allow one to determine boundary conditions without calculation.

One idea is that Feynman integrals are multi-valued functions, and not all of the singularities
present in the differential equations can appear on the first sheet of the functions.
For example, for planar integrals, there cannot be a $u$-channel singularity or branch cut,
and this information was used in ref. \cite{Henn:2013tua} to determine all non-trivial boundary
constants (the only integrals that had to be computed were integrals that trivially evaluated
to Gamma functions). 
We have seen this in the example considered in section \ref{section_intro}.
In \cite{Henn:2013nsa} it was shown that the same idea can 
be used for non-planar integrals, at the cost of introducing temporarily one additional scale.

The idea of introducing an additional variable, together with the idea of ``bootstrapping'' the 
boundary information was further used in \cite{Henn:2013nsa} to compute single-scale integrals.
Here the crucial observation was that the differential equations provide the exact form of
the answer, e.g. $x^{\eps a_{0}} \vec{f}_{0}(\eps)$ in the $x \to 0$ limit, valid for any $\eps$.
This is important since the matrix exponential contains in general terms for which the $x\to 0$ and the $\eps \to 0$ limit do not commute. The knowledge of this exact term allows one to translate between the two orders of the limits. Moreover, terms with different scaling behavior can be cleanly separated.
We give a pedagogical example of this method in section \ref{section_drinfeld}.
We note that very similar ideas were also recently applied in \cite{Dulat:2014mda}.

In the next subsections we show various examples of alphabets appearing in practice,
and comment on different ways of representing the answer.

\subsection{Examples of function alphabets}
Let us give some examples of alphabets that appear in practice in the computation
of Feynman integrals:
\begin{itemize}
\item For on-shell massless four-point integrals to three loops \cite{Gehrmann:1999as,Henn:2013tua,Henn:2013nsa}  only the letters $\{x,1+x\}$ are needed.
\item For one-variable functions, the alphabet $\{ x, 1-x,1+x\}$ often makes an appearance.
This is the case e.g. for vector boson fusion via a top quark loop to two loops \cite{Anastasiou:2006hc}
(at three loops, additional letters appear), 
Wilson line integrals \cite{Henn:2013wfa,Grozin:2014hna} forming a cusp to three loops, 
and certain contributions to Higgs cross sections at N${}^{3}$LO \cite{Hoschele:2014qsa,Dulat:2014mda}.
\item For one-loop Bhabha scattering $\{ x,1\pm x, y ,1\pm y, x+y, 1+x y\}$; however, at two loops a much larger alphabet is required \cite{Henn:2013woa}.
\item For hexagon functions in $\mathcal{N}=4$ SYM, the alphabet is  
$\{ x,y,z,1-x,1-y,1-z,1-x y,1-x z,1-y z,1-x y z\}$, at least up to three loops
\cite{Goncharov:2010jf,Dixon:2011ng,DelDuca:2011ne,Dixon:2011nj,CaronHuot:2011kk}.
\end{itemize}
Cases with several mass scales typically have a larger alphabet of the order of $10$ to $20$ letters, see e.g. \cite{vonManteuffel:2013uoa,Caola:2014lpa,Gehrmann:2014bfa,Caron-Huot:2014lda,Bell:2014zya}.
In general, there can be also polynomials of higher degree, e.g. $1+x^3$ \cite{Broadhurst:1998rz}, or even algebraic dependence on the kinematic variables, as already mentioned in one example.

It should be stressed that the statements about the alphabets occurring in certain
types of integrals are only firmly established up to a given loop order,
and in general the alphabet is not stable when increasing the loop order,
i.e. additional letters appear, or one may even leave the space of iterated integrals.

Many cases are related to a sphere with $n$ marked points \cite{Brown1}.
It is in general a difficult problem to know whether two alphabets are related
because of the freedom to redefine variables.

It is interesting to note that some alphabets have been observed to be 
related to cluster algebras \cite{ArkaniHamed:2012nw,Golden:2013xva,Golden:2014xqa}.

\subsection{Iterated integrals and hyperlogarithms}
The solution to eq. (\ref{canonicalform1}) is given by Chen iterated integrals \cite{Chen}.
In the case where the alphabet can be written in terms of rational functions (in at least one variable),
one can write the answer in terms of Goncharov polylogarithms (also called hyperlogarithms).
These functions go back to Lappo-Danilevsky \cite{bookLappo}, who introduced them precisely as solutions
to differential equations of the kind that we are discussing. 
The Goncharov polylogarithms can be defined iteratively as follows, 
\begin{align}
G(a_1,\ldots a_n ; z) = \int_0^z \frac{dt}{t-a_{1}} G(a_{2}, \ldots ,a_{n}; t) \,,
\end{align}
with 
\begin{align}
G(a_1 ;z) = \int_0^z \frac{dt}{t-a_{1}}  \,, \qquad a_{1} \neq 0\,.
\end{align}
For $a_{1}=0$, we have $G(\vec{0}_{n};z) = 1/n! \log^n(z)$.

See refs. \cite{Goncharov1,Brown1} and the lecture notes \cite{Brown2,Duhr:2014woa} for a discussion of their properties.

For a given class of Feynman integrals, one will need only a subset of allowed indices $a_i$.
One case that appears particular often in practice is that corresponding to the alphabet
$\{x,1-x,1+x\}$, i.e. indices drawn from $0, \pm 1$. This case is referred to as harmonic
polylogarithms (HPL) in the physics literature \cite{Remiddi:1999ew,Gehrmann:2001pz}.
They are denoted by $H(a_{1}, \ldots a_{n};x)$.\footnote{Note a conventional sign change - for each index equal to $1$, one needs to multiply by $-1$ to convert from the $G$ to $H$ notation.}
The example considered in section \ref{section_intro} can be solved, to any order in $\eps$, in terms of these functions.

\subsection{Different representations of the answer}
There are various ways in which the answer to the differential equations can be written.
As already discussed, it is always possible to write the answer in terms of Chen iterated integrals.
If the function alphabet is rational in at least one variable, the latter can be represented in terms of Goncharov polylogarithms. 

In practice, we are often not interested in the whole 
$\eps$ expansion, but can truncate the expansion at a given order. For example, for calculations
at next-to-next-to leading order in perturbation theory, we are typically interested in the answer only
up to weight four. In such a case, one can try to rewrite the answer in terms of a minimal function basis.
According to a conjecture by Goncharov, all weight four functions can be written in a basis
spanned by the following functions (for certain arguments, to be determined),
\begin{align}\label{degree4basis}
\{ \log x \log y \log z \log w, \log x \log y {\rm Li}_{2}(z), {\rm Li}_{2}(x) {\rm Li}_{2}(y) ,
\log x {\rm Li}_{3}(y), {\rm Li}_{4}(x), {\rm Li}_{2,2}(x,y) \}\,,
\end{align}
where 
\begin{align}
 {\rm Li}_{2,2}(x,y) = \sum_{a_1 > a_2 \ge 1 } \frac{x^{a_1} y^{a_2} }{a_1^2 a_{2}^2}  =  - \int_0^1 \frac{x \, dt}{1- x t} \log t \, {\rm Li}_{2}(x y t) \,.
\end{align}
The rewriting in terms of a minimal function basis 
can be done with the help of projections onto different terms in (\ref{degree4basis}).
Moreover, the necessary arguments in that equation can be found in a systematic way.
For these purposes, the notion of the ``symbol'' of iterated integrals is useful 
\cite{Goncharov:2010jf,Duhr:2011zq}.
We note that in the differential approach, the symbol of the answer is entirely manifest,
as it is encoded in the matrix $\tilde{A}$ of eq. (\ref{canonicalform1}).

We will illustrate the different representations 
using the example of a one-loop on-shell box integral \cite{Caron-Huot:2014lda},
called $g_6$ there, which depends on the variables $u=4 m^2/s$ and $v=4 m^2/t$.
This integral is a weight $2$ function.
As already discussed, it is always possible to write the answer in terms of Chen iterated integrals,
in this case,
\begin{align}\label{I1exactfunction}
g_6 = \int_{{\gamma}} d \, \log   \frac{ \bu -1}{\bu +1} \,  d \log  \frac{ \buv - \bu}{\buv+\bu}
+ \int_{{\gamma}} d \, \log \frac{ \bv -1}{\bv +1} \, d \, \log  \frac{ \buv - \bv}{\buv+\bv}  \,,
\end{align}
where 
\begin{align}
\beta_u = \sqrt{1+u} \,,\qquad \beta_v = \sqrt{1+v} \,,\qquad \beta_{uv} = \sqrt{1+u+v} \,.
\end{align}
The boundary condition is that $g_6$ vanishes as $u,v \to \infty$. 
The monodromy invariance allows many choices of $\gamma$, but a particularly
simple parametrization is $(u(t),v(t)) =(u/t,v/t)$, with $t \in [0,1]$.

In oder find a representation in terms of Goncharov polylogarithms,
we first need to find a change of variables that rationalizes the function alphabet.
This is achieved by setting
\begin{align}\label{wzparam}
u= \frac{(1-w^2)(1-z^2)}{(w-z)^2} \,,\qquad 
v = \frac{4 w z}{(w-z)^2} \,,
\end{align}
which allows us to write
\begin{align}\label{g6exact}
g_{6}=& -G_{-1,0}(w)+G_{0,-1}(w)-G_{0,1}(w)+G_{1,0}(w)+H_{-1,0}(z)-H_{0,-1}(z)-H_{0,1}(z) \nonumber \\
&+H_{1,0}(z)-G_0(
   w) H_{-1}(z)+G_{-1}(w) H_0(z)-G_1(w) H_0(z)-G_0(w) H_1(z) \,.
\end{align}
(This particular case only requires Goncharov polylogarithms with indices $0,\pm 1$, so they could be replaced by HPL.)

Finally, we can also write the answer in terms of the weight two minimal function basis,
namely $\{ \log x \log y, {\rm Li}_{2}(x)\}$, for certain arguments. In the present case, we have \cite{Davydychev:1993ut}
\begin{align}\label{I1exact}
g_6=& 
2 \log^2\left( \frac{\buv + \bu}{\buv + \bv} \right) + 
\log\left( \frac{\buv - \bu}{\buv + \bu} \right)  \log \left( \frac{\buv - \bv}{\buv + \bv} \right)
- \frac{\pi^2}{2} \nonumber\\
& + \sum_{i=1,2} \Big[ 
2 \, \Li_{2} \left( \frac{\beta_{i} - 1}{\buv + \beta_{i}} \right)
-2 \, \Li_{2} \left(- \frac{\buv - \beta_{i}}{\beta_{i} + 1} \right)
- \, \log^2 \left(\frac{\beta_{i} +1}{\buv + \beta_{i}} \right)
\Big]
\,.
\end{align}
Let us discuss advantages and disadvantages of the various representations.
\begin{itemize}
\item The Chen iterated integral formulation is usually the most compact way of writing the answer.
Its monodromy invariance makes it very flexible, which is of great use e.g. when computing limits, or for analytic continuation. 
Examples for numerical integration can be found in \cite{Caron-Huot:2014lda}.
Finally, we stress that the ``symbol'' of the answer is completely
manifest.
\item The Goncharov polylogarithm version of the answer amounts to fixing a specific integration contour. This usually leads to a proliferation in the number of terms, and obviously the monodromy invariance is no longer manifest, which makes e.g. analytic continuation more difficult. It should also be noted that this representation is by no means unique, since e.g. different choices of integration contour lead in general to different expressions in terms of Goncharov polylogarithm. On the positive side,
there exist dedicated numerical integration routines for these functions \cite{Bauer:2000cp,Vollinga:2004sn}.
\item The representation in terms of a minimal function basis also usually leads to a larger number of terms (compared to the Goncharov polylogarithms, one trades simple arguments of the functions for simple indices); the representation is not unique due to identities of functions involving different arguments, so that it is difficult to find an ``optimal'' representation; However, since the functions 
involved are well studied, this representation may be advantageous for numerical evaluation.
Finally, in some cases, e.g. when expressing a final result that is expected to have simple properties,
this might be made manifest by an astute choice of function arguments.
\end{itemize}

\section{Finding an optimal basis using d-log forms and generalized unitarity cuts}
\label{section_leading_singularities}

In the previous section we discussed an algebraic approach
to simplifying the system of differential equations. 
Here we discuss another (in our opinion, more natural) approach that is based on exploiting
the original integral representation, and d-log representations.

The canonical form of the differential equations in section
\ref{section_canonical} lead to answers that have two important properties:
\begin{itemize}
\item They are given by iterated integrals of uniform weight, i.e. where the $k$-th coefficient
in the $\eps$-expansion has weight $k$. 
\item No rational or algebraic factors appear, i.e. one only has  $\mathbb{Q}$-linear combinations
of iterated integrals.
\end{itemize}

Many examples of individual Feynman integrals with these properties (at least, up to some order in the $\eps$ expansion) were previously known in ${\mathcal N}=4$ SYM.
However, even understanding the transcendental weight of the leading term in the $\eps$ expansion of a given integral is not immediately obvious.
For example, at the one-loop order it is known that four-dimensional Feynman integrals give at most weight two functions.
However, in momentum space the number of initial integrations is higher (four),
and likewise in the Feynman representation, the number of integrations
is proportional to the number of propagators minus one, so that in general
neither of these representations makes the weight properties of the answer manifest.

\subsection{d-log representations}

Sometimes one can bring the integral representation to a form where the weight properties
of the answer are manifest. This is very desirable, since such an understanding usually implies also an algorithmic way of computing the answer.

\begin{figure}[t]{
\begin{center}
\includegraphics[width=0.50\textwidth]{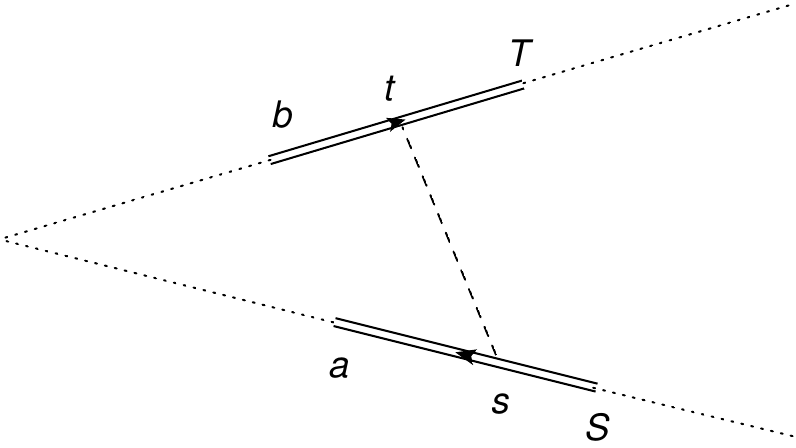}
\end{center}
\caption{Wilson line integral admitting a d-log representation. Fig. from ref. \cite{Henn:2013wfa}.}
\label{fig:example1loop}
}
\end{figure}

Let us discuss as an example the Wilson line integral shown in Fig.~\ref{fig:example1loop}, following \cite{Henn:2013wfa}.
(This is also relevant to scattering integrals, since the latter are in some cases dual to Wilson line integrals.)
 The integral over the line parameters $s$ and $t$ can be written as
\begin{align}\label{dlogform}
\int_{\Lambda}  \frac{ds \wedge dt}{s^2 + t^2 + s t (x+1/x) } = \frac{x}{1-x^2} \, \int_{\Lambda} d \log ( s+ t x)  \wedge d \log (t + s x) \,,
\end{align}
where on the r.h.s. we have dropped differentials involving $dx$ because they do not contribute to the integral,
and where the integration region ${\Lambda}$ is $s \in [a,S]$ and $t \in [b,T]$. 
In the Wilson line language, the weight properties are easy to see:
at one loop, one has two line integrations to carry out, and the result is a weight two function, as expected .

This is a first example of a d-log representation.
It has the virtue of making the fact that the answer is a $\mathbb{Q}$-linear combination
of weight $k$ functions manifest, where $k$ is the number of integrations.
This can be seen algorithmically, and the same algorithm in fact computes the function \cite{Henn:2013wfa}.
We comment that additional factors, such as $X^\eps$, that can occur in dimensional
regularization, do not change this conclusion.

Sometimes one can find similar d-log representations by manipulating the Feynman representation.
Let us take a one-loop $s$-channel bubble integral as an example, with integral mass $m$.
For general powers $a_1, a_2$ of the propagators, it is given by
\begin{align}\label{bubble1loopmassive}
\frac{\Gamma(a_1+a_2-D/2)}{\Gamma(a_1) \Gamma(a_2)}  \int_0^1 d\alpha_1 d\alpha_2 \frac{ \delta(\alpha_1+\alpha_2 -1) \alpha_1^{a_1-1} \alpha_2^{a_2-1}}{(-s \alpha_1 \alpha_2 + m^2)^{a_1+a_2-D/2}} \,.
\end{align}
We see that for $a_1 =1 , a_2 =2$ and $D=4-2 \eps$, we have a d-log representation (up to some normalization factor), with some additional, but inconsequential factor $(\ldots)^\eps$.
Inspecting further the Gamma functions, one can easily prove that upon including a normalization factor, e.g. $e^{\eps \gamma_{\rm E}}$, this yields a uniform weight one function.

This method of identifying uniform weight functions works in practice in many cases, especially for cases with bubble and/or triangle (sub)integrals.

\subsection{Generalized cuts}

In the previous section we discussed ideas to prove that a given integral has the desired uniform weight properties discussed above. In practice, sometimes a necessary condition can also be very valuable.

The idea we want to pursue is to consider discontinuities across branch cuts of the functions under consideration.
It is rather obvious that a uniform weight function will yield another uniform weight function under this operation. On the other hand, discontinuities of Feynman integrals are sometimes easier to compute than the integrals themselves.
Consider the one-loop on-shell box integral for example.
If we apply both an $s$- and $t$-channel discontinuity, we can replace its four propagators
by four delta functions. 
This localizes the four-dimensional integration. Computing the Jacobian, (and ignoring the $\eps$-dimensional part of the integration), we obtain $1/s/t$ as the answer for the cut integral.
This explains the normalization factor $s t$ in eq. (\ref{eqbasisbox}) of the uniform weight basis discussed there.

We now generalize the above idea from unitarity cuts to generalized cuts, where any number
of propagators can be cut (replaced by delta functions). 
Such integrals satisfy the same IBP relations and hence the same differential equations as the original functions, albeit with different boundary conditions.
Roughly speaking, thinking about
a system of DE, each way of choosing cuts projects onto a smaller, but nontrivial subsystem.
If we are looking for a canonical form of the full system, the same canonical form has
to be present also at the level of the generalized cuts.

Therefore, we can impose as a criterion for a putative basis of uniform weight functions that
any generalized cut (i.e. replacing any number of propagators by delta functions) should give
a pure uniform weight function.

This is a very powerful test that can be easily performed. 
In particular, as we will see, at higher loops one can ``recycle'' the knowledge obtained previously from lower loops.
Let us give two examples.

{\bf Example: One-loop pentagon integral.} \\
Consider the one-loop pentagon integral with some numerator $N(k)$, 
\begin{align}
\int \frac{d^{D}k}{i \pi^{D/2}} \frac{N(k)}{k^2 (k+p_1)^2 (k+p_1 + p_2)^2 (k+p_1+p_2+p_3)^2 (k-p_5)^2 } \,.
\end{align}
There are five possibilities for computing a maximal cut (cutting four propagators).
The answer on a given cut will be of the form
\begin{align}
\frac{N(k_{*})}{J P(k_*)}\,,
\end{align}
where $k_{*}$ is a solution to the cut conditions, $J$ is the Jacobian, and $P$ is the uncut propagator factor, e.g. $P(k)=(k-p_5)^2$, if the first four propagators are cut.
One may verify that for a constant (i.e. $k$-independent) $N$, there is no choice of $N$ such that one obtains a $\mathbb{Q}$ number on each cut.
Therefore the scalar pentagon integral in four dimensions is not a pure uniform weight integral.
We prefer not to use it in our basis.

However, it is now clear how to proceed. We can allow a more general numerator $N(k)$, and impose the condition that all cuts should yield numbers in  $\mathbb{Q}$ only.
We refer to reference \cite{ArkaniHamed:2010gh} where this philosophy was introduced and
many integrals with these properties were constructed.
This reference also gives an introduction to momentum twistors that allow one to visualize more easily the geometry of the cuts, and in terms of which the cut solutions are simple.

{\bf Example: Two-loop planar and non-planar box integrals.} \\
For a multi-loop example, we may consider a massless double box integral.
Here we wish to illustrate how the knowledge of lower-loop information can be recycled.
Consider an ansatz of the form
\begin{align}
\int \frac{d^D k_1 d^D k_2}{(i \pi^{D/2})^2} \frac{N(k_1)}{k_1^2 (k_1+p_1)^2 (k_1+p_{12})^2 k_2^2 (k_2+p_{12})^2 (k_2-p_4)^2 (k_1-k_2)^2}  \,,
\end{align}
where $p_{12}=p_1+p_2$.
Consider cutting all propagators of the $k_2$ integral, as well as two adjacent propagators.
Localizing the $k_2$ integration we obtain a Jacobian factor of $1/s/(k_1-p_{4})^2$, so that we are left with (ignoring the $\eps$-dimensional part of the integration)
\begin{align}
\frac{1}{s} \int \frac{d^D k_1}{(i \pi^{D/2})} \frac{N(k_1)}{k_1^2 (k_1+p_1)^2 (k_1+p_{12})^2 (k_1-p_4)^2 }  \,.
\end{align} 
We recognize this as a one-loop box integral, with a numerator $N(k_1)$.
Our knowledge of one-loop integrals now tells us that choosing $N(k_1)  = s^2 t $ 
makes this a uniform weight box integral.
However, there is a second possibility: choosing $N(k_1) = s^2 (k_1-p_4)^2$, we obtain
a triangle integral that is also a uniform weight function (since triangle and box integrand simply differ by the fact that for the triangle one of the points is at infinity).
This simple analysis explains the choice of integral basis for the double box integrals in ref. \cite{Henn:2013pwa}.

For a non-planar example, consider the two-loop non-planar double box integral.
It contains both a pentagon and a box subintegral. Let us consider the generalized cut that localizes the box subintegral. The Jacobian $1/(P_1 P_2 )$ gives two propagator factors that depend on the remaining loop momentum. This means that the cut integral contains a scalar pentagon integral.
We have seen above that this integral does not have the desired uniform weight properties
that we are looking for. However, we already know how to remedy this problem, by using as the starting point a non-planar double box with appropriate numerator factors. In complete analogy to the analysis above, suitable candidate numerator factors are $P_1$, $P_2$, or $P_1 P_2$, multiplied by an appropriate overall normalization.

\begin{figure}[ht]
\begin{center}
\includegraphics{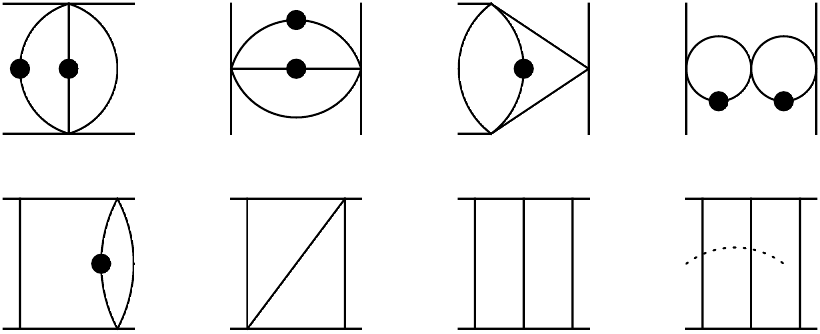}%
\end{center}
\caption{Integral basis for on-shell planar two-loop integrals. 
Dots indicate doubled propagators, and the dotted line an inverse propagator. 
Loop-momentum-independent factors are not shown. Figure from ref. \cite{Henn:2013pwa}.
\label{fig:4ptints}}
\end{figure}

{\bf Example: Two-loop $2\to2$ integrals of ref. \cite{Henn:2013pwa}.}\\
As an application to the discussion of this and the last subsection we are now in a
position to fully understand the basis choice made e.g. in ref. \cite{Henn:2013pwa}, see. Fig. (\ref{fig:4ptints}).

The propagator-type integrals can all be directly evaluated in terms of (iterations of the one-loop) formula (\ref{formula_bubble}).
It is easy to see that for $D=4-2 \eps$ dimensions, this formula gives results of homogeneous transcendental weight for the choice of doubled lines as shown in Fig.~\ref{fig:4ptints}.

Let us now discuss the remaining integrals shown in that figure.
Integrals having bubble sub-integrals are also immediate: integrating out the bubble integral
yields a Feynman integral already known from the analysis of the previous loop order \cite{Henn:2013tua}.
E.g., the first integral in the second line is related to a box integral.
The latter is known to be of uniform weight, with known normalization factor (the latter can
be determined e.g. through a maximal cut). 
Similarly, the third integral in the first line is related to a one-loop triangle. 
Triangle integrals in four dimensions have the same properties as box integrals.
This can be seen by employing momentum twistor variables \cite{Hodges:2009hk,ArkaniHamed:2010gh}, or in the embedding space
formalism \cite{embedding}. Roughly speaking, the triangle is a box with one point at infinity.

The second integral in the second line can be understood in two different ways.
One possibility is to introduce one Feynman parameter to combine two propagators
adjacent to an incoming on-shell momentum. In this way, one obtains a one-parameter
integral (with a d-logarithmic integration kernel) over an integral it is easily seen to be
of uniform weight, in complete analogy to what was discussed above.
Another possibility is to use the embedding formalism, in which case the cut analysis
as explained above is straightforward.

Finally, for the most complicated, seven-propagator sector, the basis choice 
was explained above using cuts. One numerator is just the overall normalization
$N = s^2 t$, while the other one contained a loop momentum dependence, $s^2 (k_1-p_4)^2$,
which is indicated a dashed line in the figure.

This explains how to choose a basis of uniform weight integrals for this problem.
The critical reader may wonder about the absolute weight for the individual basis elements.
The latter are indeed different: for example, the double box integrals have weight $4$,
whereas e.g. the double bubble integrals as defined here have weight $2$.
This overall difference in weight 
is compensated 
 by powers of $\eps$ included 
in the definition of the basis in ref. \cite{Henn:2013tua}; 
this final adjustment leads to the
canonical form quoted there.

We leave it to the interested reader as an exercise to reproduce the basis choice
for higher-loop integrals \cite{Henn:2013tua}, or for partially off-shell kinematics \cite{Henn:2013nsa,Caola:2014lpa}, or cases with masses \cite{Caron-Huot:2014lda}.
Further explicit examples can be found in section \ref{section_drinfeld}.

We note that this method typically gives a more compact form of the basis compared to the
algebraic method of section \ref{section_algorithm},
and it is more likely to make physically important properties, such as e.g. infrared 
behavior, manifest.

We close this section with a number of remarks.
\begin{itemize}
\item While the cut analysis in this section only provides consistency checks for an integral to be of uniform weight, we wish to emphasize that this hypothesis can be immediately tested (and proved!) by deriving the system of differential equations for it.
\item In practice, one often finds that after this analysis, the system obtained is close to the desired canonical form. It may happen that small transformations, in the spirit of section \ref{section_algorithm}, are required to fix e.g.  terms that vanish on the cuts considered.
This combined analysis is what we have found most useful in practice.
\item The geometry of the cut equations for massless on-shell Feynman integrals
is particularly transparent and simple when using momentum twistor variables \cite{Hodges:2009hk,ArkaniHamed:2010gh}.
As was already mentioned, the same variables, or also the embedding formalism \cite{embedding}, also have the virtue of making the 
point at infinity manifest, so that triangle integrals and box integrals are in fact treated on the same footing.
\item We also wish to mention that the above criteria could be tested algorithmically,
for a given graph. In practice, we have not found this necessary since the application
by hand is very simple.
\item As we have seen in the examples, the cuts relate different loop orders,
and one can therefore ``recycle'' information about an optimal basis of lower loops
when going to the next loop order; also, once the behavior of certain subgraphs
is understood, they can be used at higher loops as ``lego blocks'', i.e. without having
to revisit the justification for using them.
\item We also mentioned that cuts very naturally project the DE onto subsectors;
this allows one to gradually construct the full system of DE, dealing with smaller
matrices only at a given time. This can be of practical importance, since it means
that one has to set up the entire $N \times N$ system only when the optimal basis
is essentially known, and since the DE are much simpler in the optimal basis, this
leads to a significant speed-up in the algebraic manipulations needed.
\item Finally, it is important to point out that the initial motivation for considering 
integrals of this type is to make infrared properties of scattering amplitudes especially simple
by choosing basis integrals that are either infrared finite or ``almost'' finite.
This is related to generalized cuts since the latter can probe regions of loop integration that produce infrared divergences. It was observed that choosing integrals defined in this way made the
answers particularly simple, even before carrying out the loop integration \cite{ArkaniHamed:2010kv,Drummond:2010mb,ArkaniHamed:2010gh}.
\end{itemize}

\subsection{Momentum-space d-log representations}

We wish to mention that there exists another type of d-log representation
that is closely related to the cut properties discussed in the last section.
Namely, it was observed that certain loop integrals/integrands can be 
algebraically put into a momentum-space d-log form.

For example, for the integrand of the one-loop box integral 
\begin{align}
\mathcal{I}_4(\alpha) =  \frac{d^4k\,(p_1+p_2)^2(p_2+p_3)^2}{k^2(k+p_1)^2(k+p_{12})^2(k-p_4)^2} \,,
\end{align}
this representation is \cite{Arkani-Hamed:2014via}
\begin{align}
\mathcal{I}_4(\alpha)\!=\!d\!\log  \frac{k^2}{(k-k_*)^2} \wedge  d\!\log \frac{(k+p_1)^2}{(k-k_*)^2} \wedge d\!\log \frac{(k+p_{12})^2}{(k-k_*)^2} \wedge d\!\log \frac{(k-p_4)^2}{(k-k_*)^2}\,, 
\end{align}
where $k_*$ is one of the quadruple cuts solutions of the box.
More examples of d-log representations of this kind, for example for the planar and non-planar double box integrals discussed in the previous subsection are given in ref. \cite{Arkani-Hamed:2014via}.

One virtue of this representation is that it is straightforward to take generalized cuts,
since no Jacobians have to be computed.
See also the discussion in \cite{ArkaniHamed:2012nw} regarding the expected 
transcendental weight properties of the integrated result.

For progress in the direction of a direct integration in momentum space
starting from such expressions see refs. \cite{Lipstein:2012vs,Lipstein:2013xra}.

\section{Bootstrapping single-scale integrals using the Drinfeld associator}
\label{section_drinfeld}

 In this section we give a simple pedagogical example of the application of differential equations to single-scale Feynman integrals. We explain how to find a basis of integrals of uniform weight in practice, 
 and  discuss the structure of the differential equations, placing particular emphasis on the singular points and asymptotic limits. This is based on refs. \cite{Henn:2013pwa,Henn:2013nsa}.

Naively, one cannot use DE to compute single-scale integrals, since their
scale dependence is trivial. The main idea is to introduce a parameter $x$ in a natural way,
so that for $x=0$ the original integrals are recovered. The system of DE w.r.t. $x$ turns out to have another  singular point at $x=1$, i.e.
\begin{align}\label{eqKZ}
\partial_x \vec{f}(x,\eps) = \eps \left[ \frac{a}{x} + \frac{b}{1-x} \right] \vec{f}(x,\eps) \,.
\end{align}
The point $x=1$ turns out to yield a simple boundary condition, without computation,
similar to the example considered in section \ref{section_intro}. One can then transport
this boundary information back to $x=0$, via the DE. This is precisely what is computed
by the Drinfeld associator. The latter depends only on the matrices $a$ and $b$,
and we can compute its expansion in $\eps$, and hence the original single-scale
integrals, to any desired order in $\eps$.\footnote{The recent ref. \cite{Papadopoulos:2014hla} also uses the idea of introducing an additional parameter that is expected to have simple boundary values.}
Moreover, it follows from eq. (\ref{eqKZ}) that the associator contains only multiple zeta values.

\begin{figure}[t] 
\captionsetup[subfigure]{labelformat=empty}
\begin{center}
\subfloat[$f_7$]{\includegraphics[width=0.45\textwidth]{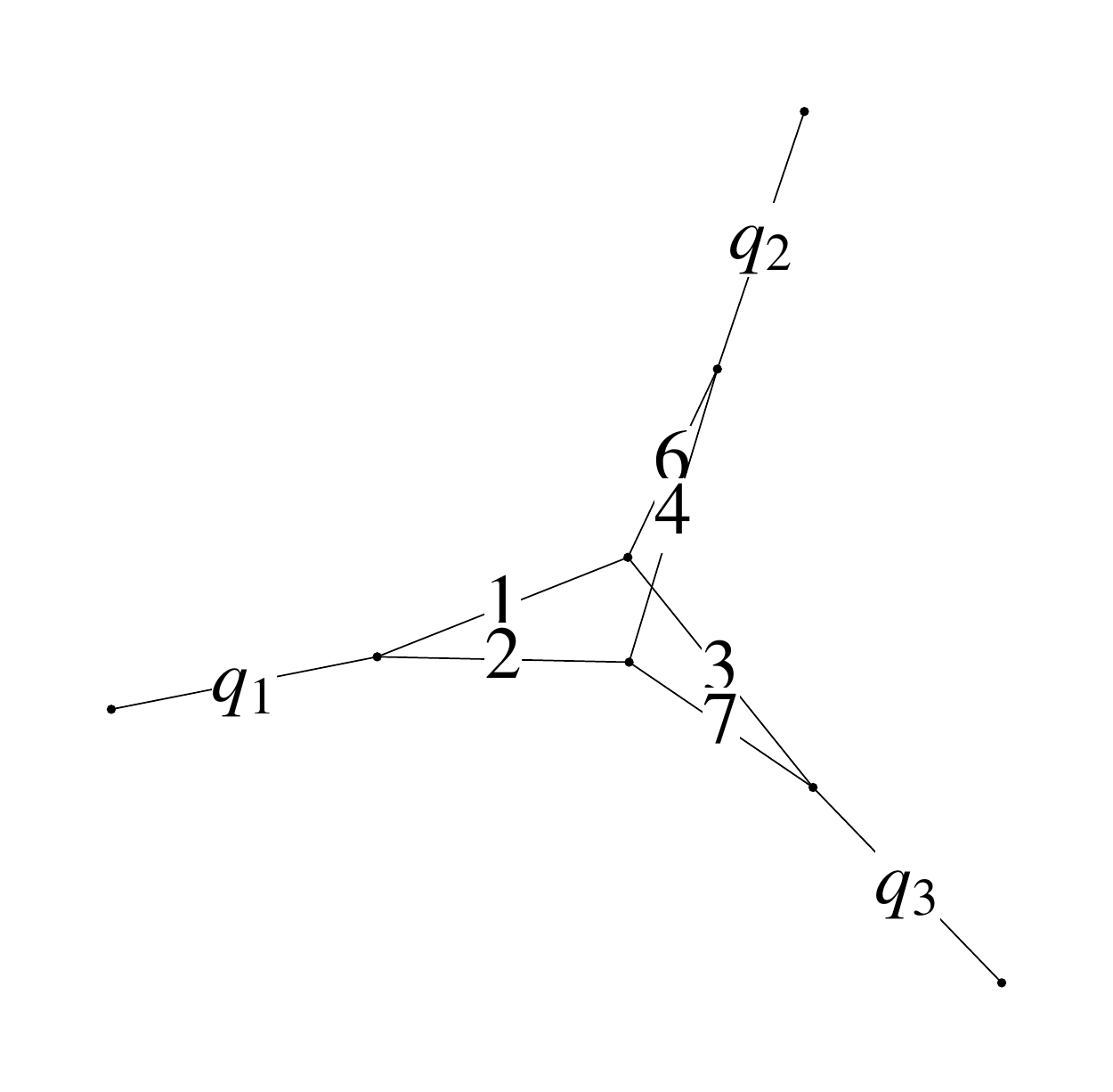}}
\caption{Non-planar  massless form factor integral. $q_1^2 \neq 0, q_{3}^2=0$, and we will consider both $q_2^2 = 0$ and $q_{2}^2 \neq 0$.}
\label{fig:NPFF}
\end{center}
\end{figure}

Consider the family of massless non-planar form factor integrals at two loops,
\begin{align}
G_{a_1, \ldots a_7} := & e^{2 \eps \gamma_{E}} \int \frac{d^{D}k_1}{i\pi^{D/2}}  
\frac{d^{D}k_2}{i \pi^{D/2}} \frac{1}{[-k_1^2]^{a_1} [-(k_1 + q_1)^2]^{a_2} [-k_2^2]^{a_{3}} }\times \nonumber \\
&  \times \frac{[-(k_1 - q_3)^2]^{-a_{5}}}{[-(k_1 - k_2 + q_1 + q_3)^2]^{-a_4 } [-(k_1 - 
    k_2)^2]^{a_6 } [-(k_2 - q_3)^2 ]^{a_7} }\,,
\end{align}
where $q^\mu_{1}+q^\mu_{2}+q^\mu_{3}=0$ and $a_{5}\le0$.
We are interested in this for the on-shell case $q_{2}^2=q_{3}^2=0$.

\begin{figure}[t] 
\captionsetup[subfigure]{labelformat=empty}
\begin{center}
\subfloat[$f_1$]{\includegraphics[width=0.35\textwidth]{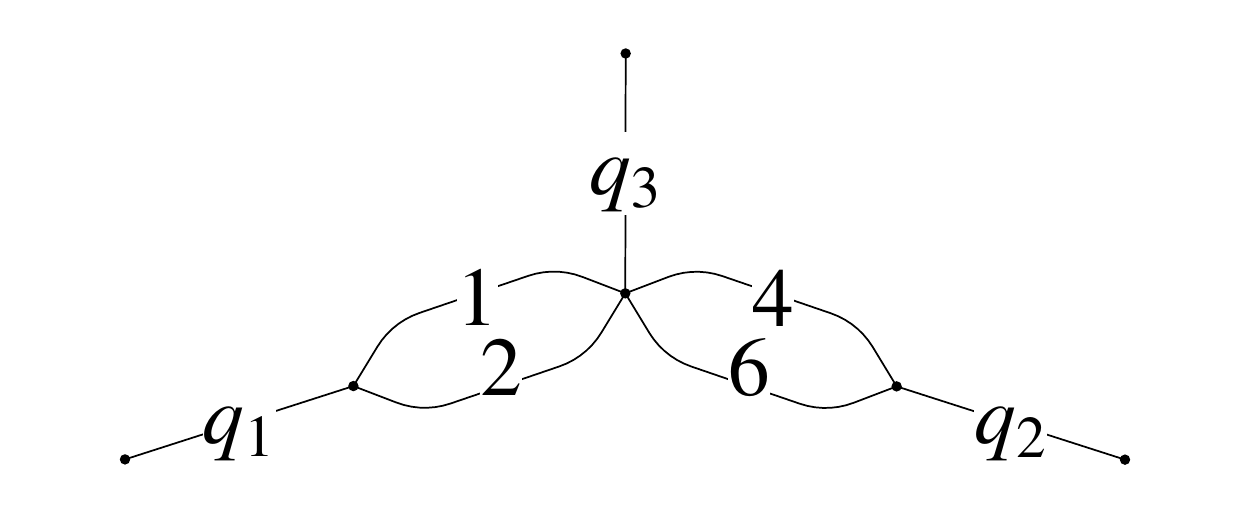}}
\subfloat[$f_2$]{\includegraphics[width=0.35\textwidth]{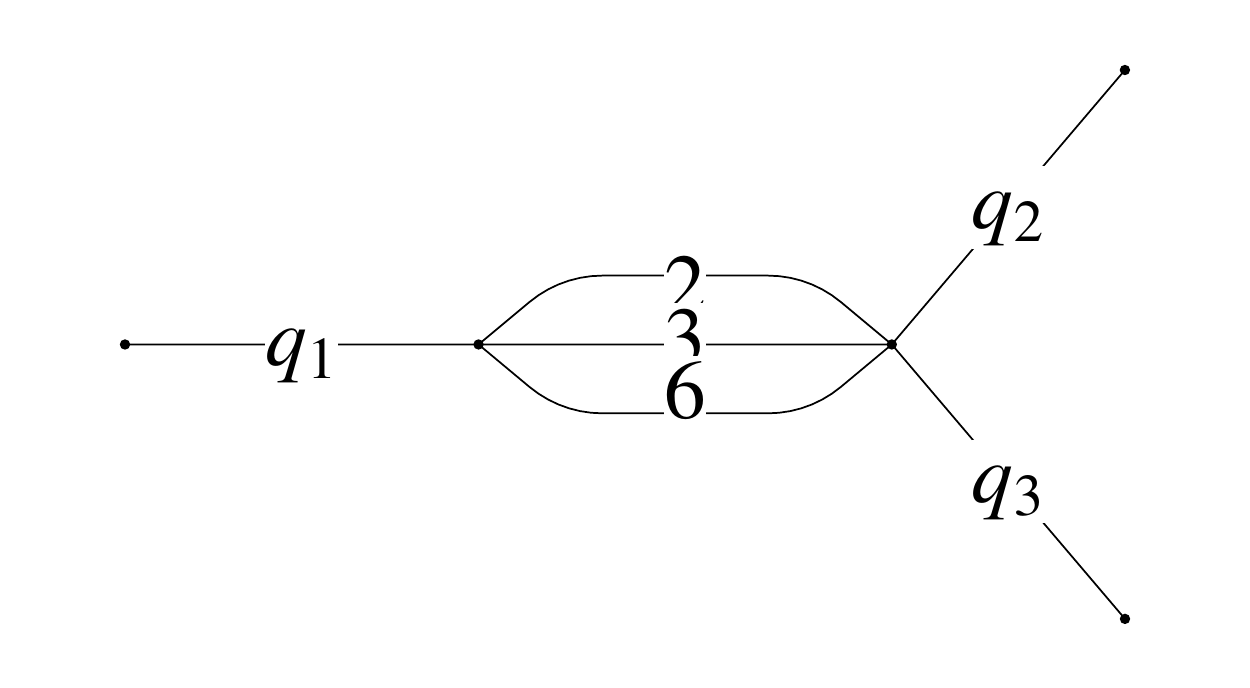}}
\subfloat[$f_3$]{\includegraphics[width=0.35\textwidth]{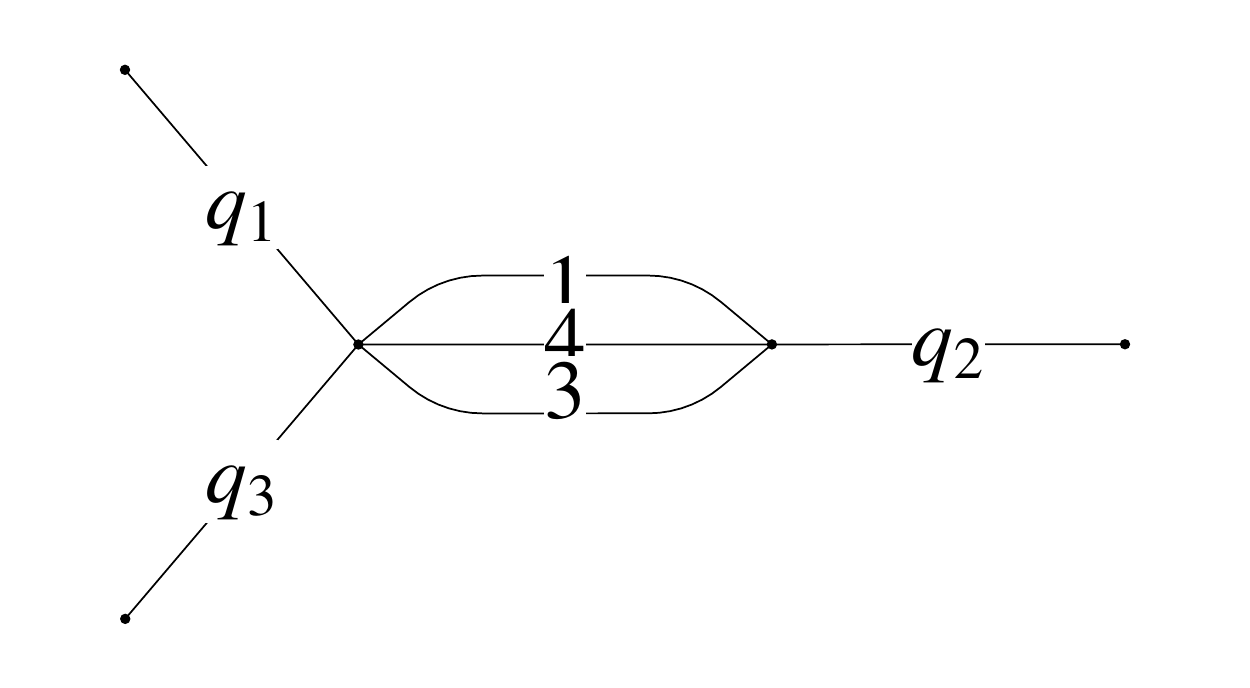}}
\newline
\subfloat[$f_4$]{\includegraphics[width=0.35\textwidth]{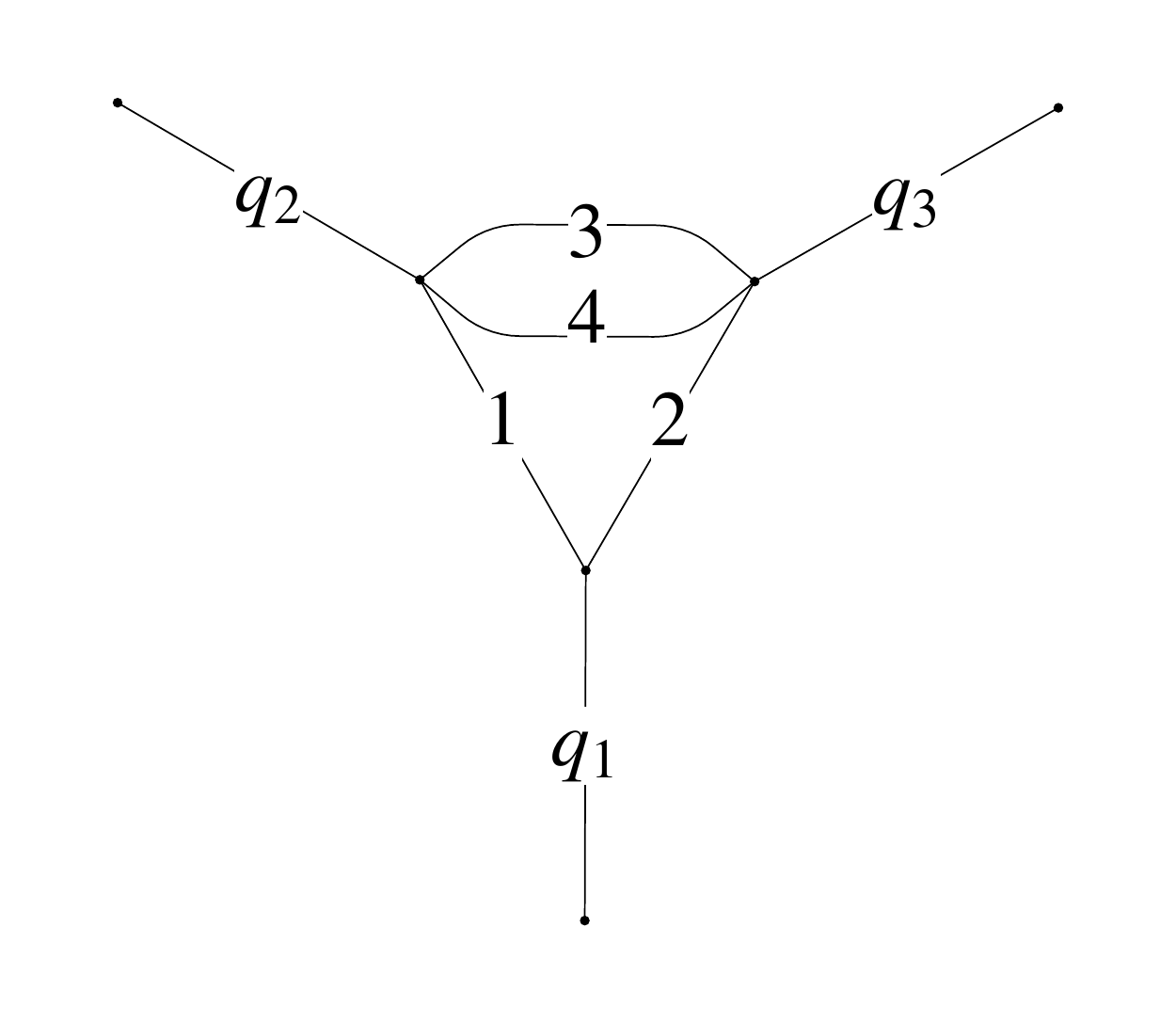}}
\subfloat[$f_5$]{\includegraphics[width=0.35\textwidth]{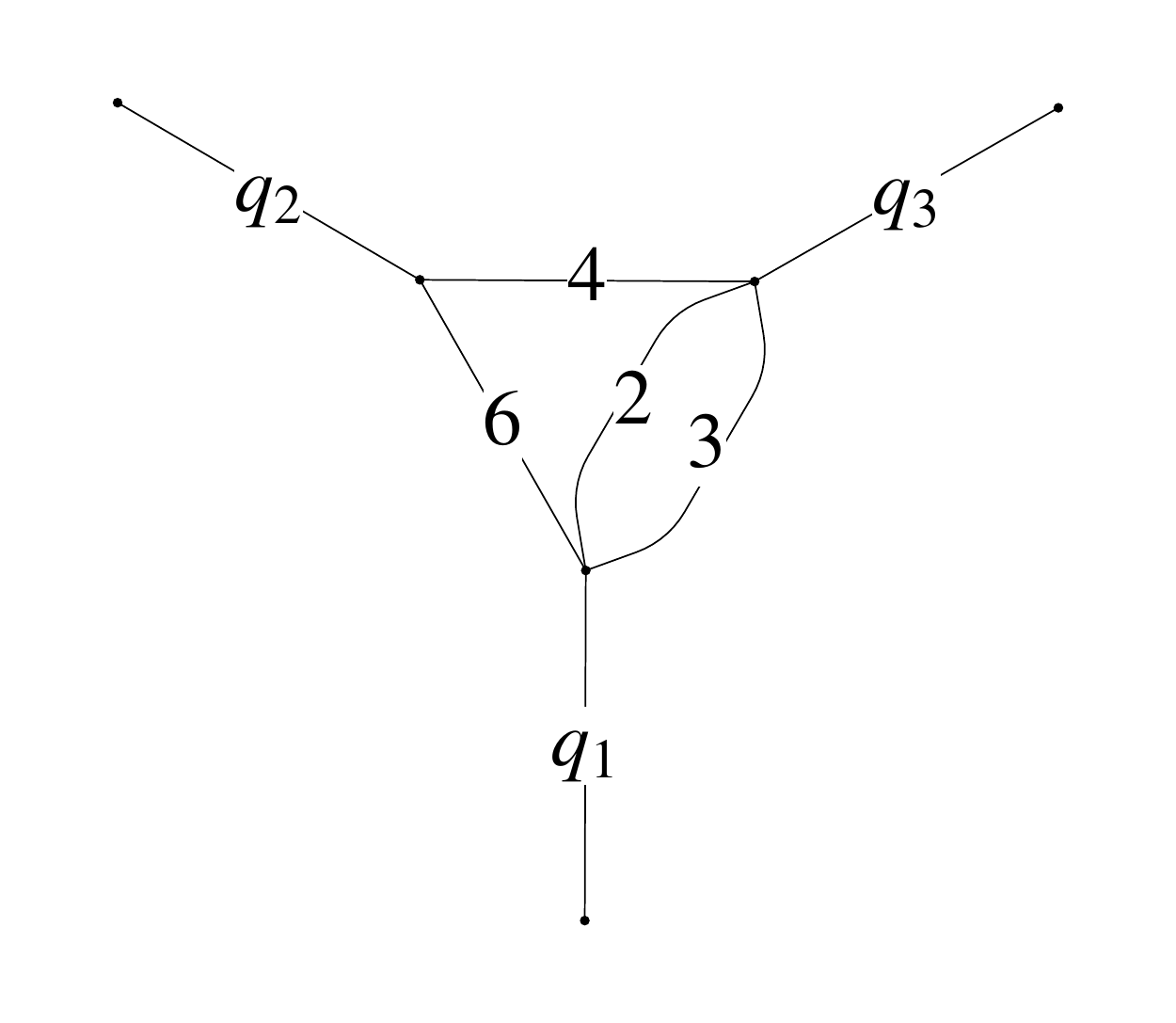}}
\subfloat[$f_6$]{\includegraphics[width=0.35\textwidth]{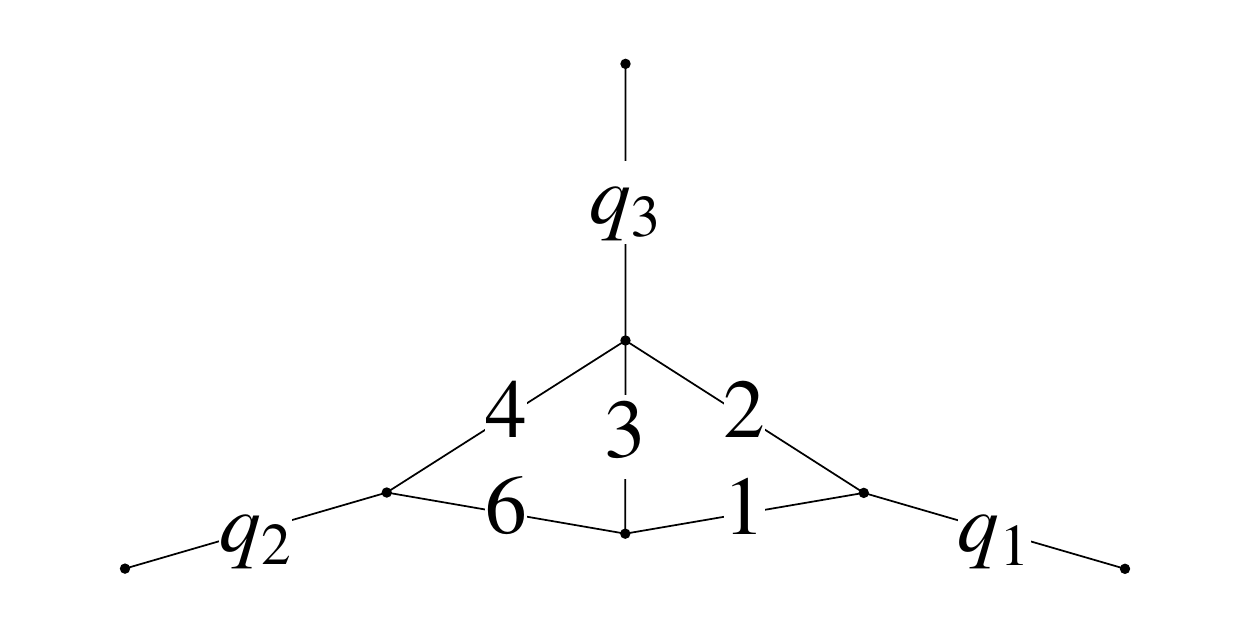}}
\caption{Basis integrals of the integral family (in addition to $f_7$).}
\label{fig:basisFF}
\end{center}
\end{figure}

There is one non-trivial crossed ladder integral, see Fig.~\ref{fig:NPFF}, that we would like to evaluate using differential equations. In order to do this, we introduce another scale by letting $q_{2}^2 \neq 0$.
In that case, there are seven basis integrals. 
They are shown in Figs.~\ref{fig:NPFF} and \ref{fig:basisFF}.
The integrals now depend on two scales, $q_1^2$ and $q_2^2$, and we can derive differential equations in $x=q_2^2/q_1^2$.

On general grounds it is known that we will obtain a first-oder system of differential equations for the basis integrals. In the following, we will explain how to simplify that system by making an appropriate choice of basis integrals.

\subsection{Choosing basis integrals that have uniform weight}
We would like to choose basis integrals that have uniform weight, i.e. that
are built from linear combinations of functions of a given weight.
We will call such functions UT (loosely for uniform transcendental degree).
There are several ideas for how to find such functions. Let us apply a few of them in the present case.

The bubble (or propagator-type integrals) in the first line of Fig.~\ref{fig:basisFF} are elementary,
as they evaluate to products of Gamma functions, for any value of the propagator exponents. 
Here we just note the choice that leads to functions with uniform weight (it will become clear presently  how we arrived at this choice).
\begin{align}
f_{1} =& \eps^2 (-q_1^2)^{2 \eps}  q_1^2 q_2^2 G_{2,1,0,2,0,1,0} \,, \\
f_{2} =& \eps^2 (-q_1^2)^{2 \eps}   q_1^2 G_{0,2,2,0,0,1,0} \,, \\
f_{3} =& \eps^2 (-q_1^2)^{2 \eps}  q_2^2  G_{2,0,2,1,0,0,0} \,, 
\end{align}
Iterating the formula for the basic one-loop integral given in eq. (\ref{formula_bubble}), we have, e.g.
\begin{align}\label{eqf1}
f_{1} =& e^{2 \eps \gamma_{E}} x^{-\eps} \Gamma^4(1 - \eps) \Gamma^2(1 + \eps)/\Gamma^2(
 1 - 2 \eps) \nonumber \\
=& x^{-\eps}  \left[ 1  - \eps^2 \frac{\pi^2}{6} - \eps^3 \frac{14}{3} \zeta_3 + \cO(\eps^4) \right]  \,,
\end{align}
and we indeed see that each term in the $\eps$ expansion has uniform weight.
This is already obvious from the first line of eq. (\ref{eqf1}), if one takes into account that
\begin{align}
\log\left[ \Gamma(1+\eps) \right] = -\gamma_{E} \eps + \sum_{k\ge 2} (-1)^k \zeta_k \frac{\eps^k}{k}   \,.
\end{align}

The situation is more interesting for the triangle integrals shown in the second line of Fig.~\ref{fig:basisFF}. Consider $f_{4}$ for example.
We already know that we can always integrate out a massless bubble insertion. The resulting integral will be a triangle integral, where one line has a shifted power $a_3+a_4-D/2$. 
We will argue that the choice $a_1=a_2=a_3=1, a_4=2$ is a good one.
This can be see e.g. from the Feynman representation, which reads
\begin{align}
e^{\eps \gamma_{E}} \frac{\Gamma(1+2 \eps)}{\Gamma(1+\eps)}
 \int  \frac{   d\alpha_1 d\alpha_2 d\alpha_3  \delta(1-\sum_i  \alpha_i) \alpha_3^{\eps}}{[\alpha_1 \alpha_2 (-q_1^2)+\alpha_1 \alpha_3(-q^2_2) ]^{1+ \eps} } \,.
\end{align}
We will see that this integral is UT, without having to evaluate it.
The main point is that its integrand is a rational form with logarithmic singularities.
Solving the $\delta$ function e.g. by $\alpha_1 = y, \alpha_2 = (1-y) z, \alpha_3 =(1-y)(1-z)$, 
we see that the $y$ integration is elementary, and the remaining integral becomes
\begin{align}
\int_0^1 \frac{dz  (1-z)^\eps }{[ z (-q_1^2) + (1-z) (-q^2_2) ]^{1+\eps }} = \frac{1}{q_2^2 -q_1^2 }  \int_0^1  d \log(X)  \,  X^{-\eps} (1-z)^\eps
\end{align}
with $X=z (-q_1^2) + (1-z) (-q^2_2)$.
From this is is clear that there is a unique normalization factor, $q_2^2 -q_1^2$. Moreover, 
upon integration, the function at $\eps=0$ will have weight $1$. It is easy to convince oneself that
expanding to $\eps^k$ will increase the weight of the function by $k$. Hence we conclude that
\begin{align}
f_{4} =  \eps^3 (-q^2_1)^{2 \eps} (-q_1^2 + q^2_2) G_{1, 1, 1, 2, 0, 0, 0, 0, 0}\,,
\end{align}
is a UT basis integral (the other factors were chosen for later convenience).
In the same way, one arrives at the choices
\begin{align}
f_{5} =& \eps^3 (-q^2_1)^{2 \eps} (-q_1^2 + q^2_2) G_{0, 2, 1, 1, 0, 1, 0, 0, 0}\,, \\
f_{6} =& \eps^4 (-q^2_1)^{2 \eps} (-q_1^2 + q^2_2) G_{1, 1, 1, 1, 0, 1, 0, 0, 0}\,.
\end{align}

Finally, for the last basis integral, it is more convenient to use a different
approach based on unitarity cuts. The idea is that, if an integral is UT, this property is preserved by unitarity cuts. In practice, it is convenient to perform maximum cuts that localize the loop momentum (in four dimensions). We have already seen in the above analysis that often ``$\cO(\eps)$'' terms do not influence the UT property, and therefore we will ignore them in the first approximation. 
(If they do matter, one can always perform a more careful analysis.)
In the present situation, we can perform a maximal cut.
We proceed in two steps: cutting all visible propagators allows us to localize, say the $k_{2}$ integration. The resulting Jacobian produces further propagator factors, which we cut as well.
In this way, we obtain a leading singularity of $1/(-q^2_1 + q_2^2)^2$. We are thus led to define the candidate integral
\begin{align}
f_{7} = \eps^4 (-q^2_1)^{2 \eps} (-q_1^2 + q_2^2)^2 G_{1, 1, 1, 1, 0, 1, 1, 0, 0} \,.
\end{align}

A comment is in order here. Unlike the analysis of the Feynman parametrizations, the cut analysis
did not lead to a proof that the integral in question is UT. However, we will be able to prove this presently from the differential equations. The leading singularity method is invaluable for finding candidate integrals that can then be rigorously proven to be UT.

\subsection{Differential equations, boundary conditions, and solution}

The candidate integrals $\vec{f} = \{ f_1 , \ldots , f_7 \}$ chosen above are all dimensionless functions of the single variable $x=q_{2}^2/q_{1}^2$, and $\eps$. 
We can derive differential equations in $x$ by differentiating the Feynman integrals w.r.t. $q_1$ or $q_2$, and using integration-by-parts identities to re-express the result of the differentiation in terms of  $\vec{f}$.

In this way, we find the following system of differential equations (DE),
\begin{align}\label{DE1}
\partial_x  \vec{f}(x,\eps)  = \eps \left[ \frac{a}{x} + \frac{b}{1-x} \right] \vec{f}(x,\eps) \,,
\end{align}
with
\begin{align}
a={\small \left(
\begin{array}{ccccccc}
 -1 & 0 & 0 & 0 & 0 & 0 & 0 \\
 0 & 0 & 0 & 0 & 0 & 0 & 0 \\
 0 & 0 & -2 & 0 & 0 & 0 & 0 \\
 0 & 0 & \frac{1}{2} & 0 & 0 & 0 & 0 \\
 0 & \frac{1}{2} & 0 & 0 & -1 & 0 & 0 \\
 -1 & -\frac{1}{2} & -\frac{1}{2} & -2 & 2 & -2 & 0 \\
 0 & -3 & 3 & 4 & 4 & 4 & 0
\end{array}
\right) }\,, \quad
b={\small \left(
\begin{array}{ccccccc}
 0 & 0 & 0 & 0 & 0 & 0 & 0 \\
 0 & 0 & 0 & 0 & 0 & 0 & 0 \\
 0 & 0 & 0 & 0 & 0 & 0 & 0 \\
 0 & 0 & 0 & 1 & 0 & 0 & 0 \\
 0 & 0 & 0 & 0 & 1 & 0 & 0 \\
 0 & 0 & 0 & 0 & 0 & -2 & 0 \\
 0 & 0 & 0 & 0 & 0 & 0 & 2
\end{array}
\right)}\,.
\end{align}
We see that the DE have three singular points, $x=0,1,\infty$.
They correspond to different physical limits, namely $q_2^2 =0, q_3 \to 0, q_1^2 = 0$.
At this point, it is useful to recall the analytic structure of the functions under consideration.
From the Feynman parametrization it is obvious that they are real in the unphysical region $q_1^2 <0, q_2^2 <0$, i.e. for any $x>0$. As a consequence, they cannot have a singularity or branch cut at $x=1$. We can use this information as a boundary condition for the integrals. In fact, thanks to their normalization factors, the integrals $f_{4}, f_{5}, f_{6}, f_{7}$ have to vanish at $x=1$. Since $f_{1}, f_{2}, f_{3}$ are elementary, this provides the complete boundary information, without any calculation.

We are now in the position to prove that the functions $\vec{f}$ are indeed UT, as expected.
In fact, this is obvious: expanding them in $\eps$, according to
\begin{align}
\vec{f}(x,\eps) = \sum_{k \ge 0} \eps^k \vec{f}^{(k)}(x) \,,
\end{align}
we see that the DE decouples at each order in $\eps$. It follows that the solution at order $k$ will be a $k$-fold iterated integral (over rational differential forms with logarithmic singularities), and hence has uniform weight. One could worry that the boundary constants might spoil the UT property, but it is easy to see that this is not the case.

We can then write the solution to eq. (\ref{DE1}) in terms of harmonic polylogarithms, 
to any order in $\eps$. For example, to order $\eps^2$,
\begin{align}\label{result1}
f_1=& 1-\eps
   H_0(x)+ \eps^2 \left(H_{0,0}(x)-\frac{\pi ^2}{6}\right) + \cO(\eps^3)\,, \\   
   f_2=& -1 +\eps^2  \frac{\pi ^2 }{6} + \cO(\eps^3) \,,\\
   f_3=& -1   +2 \eps H_0(x)+   \eps^2 \left(\frac{\pi
   ^2}{6}-4 H_{0,0}(x)\right) + \cO(\eps^3)\,,\\
   f_4=&-\eps\frac{1}{2}
    H_0(x) +  \eps^2
   \left(H_{0,0}(x)-\frac{1}{2} H_{1,0}(x)-\frac{\pi ^2}{12}\right) + \cO(\eps^3) \,,\\
   f_5=&-\frac{1}{2} \eps H_0(x)  + \eps^2 \left(\frac{1}{2} H_{0,0}(x)-\frac{1}{2}
   H_{1,0}(x)-\frac{\pi ^2}{12}\right) + \cO(\eps^3) \,,\\
   f_6=& \cO(\eps^3) \,,\\
   f_7=&2
   \eps^2 H_{0,0}(x)+ \cO(\eps^3)  \,.
\end{align}
As expected from the discussion above, $f_{4}$ starts with a term weight one.
Of course, it is straightforward to expand to any desired order in $\eps$.

\subsection{Asymptotic limits, matching, and result for single-scale integral}

We would like to recover the single-scale functions at $q_{2}^2=  0$. This limit does not in general commute with the $\eps\to0$ limit. Therefore, we have to analyze it in more detail.
Eq. (\ref{DE1}) provides the information we need. We can solve it for small but finite $x$, at any value of $\eps$. The leading term is given by the matrix exponential
\begin{align}\label{DEasy}
\vec{f}(x,\eps) \sim x^{\eps a} \vec{g}(\eps) \,,
\end{align}
with $\vec{g}(\eps)$ the boundary information.
Hence the eigenvalues of $a$ determine the asymptotic behavior of the functions.
We find that they are $0,-1,-2$. The presence of non-zero eigenvalues indicates that the limits $x\to0$ and $\eps \to 0$ do not commute. 
However, eq. (\ref{DEasy}) gives us enough information to fix this problem.
Indeed, we can start with eq. (\ref{DEasy}) and send $\eps \to 0$, for small but non-zero $x$.
This allows us to make contact with the result of the previous section, where we determined the perturbative expansion in $\eps$.
This information allows us to determine the `matching coefficients' $\vec{g}(\eps)$, in the $\eps$ expansion.
Having this determined the boundary constants in eq. (\ref{DEasy}), we can use it to take $x\to 0$ for finite $\eps$. This is the limit that takes us to the single-scale function we are interested in. 
We obtain it now by keeping only the terms corresponding to eigenvalues $0$.

Going throughout the steps described above, we find, to order $\eps^6$,
\begin{align}
f_{7}(0,\eps) =& 1- \eps^2 \pi^2  - \frac{83}{3} \zeta_3 \eps^3 - \frac{59}{120} \pi^4 \eps^4 + \eps^5 \left( \frac{79}{6} \pi^2 \zeta_3 -\frac{587}{5} \zeta_5 \right) \nonumber \\
& + \eps^6 \left( \frac{2567}{9} \zeta_3^2 + \frac{59}{1512} \pi^6 \right)  + \cO(\eps^7) \,.
\end{align}
This is in perfect agreement with results found in the literature. It is straightforward computer algebra to extend this to higher orders in the $\eps$ expansion.

\section{Discussion and conclusion}

We have given a review of the differential equations technique for the
computation of Feynman integrals.
We pointed out general properties of the equations that allow one
to make the singularity structure manifest. 
Moreover, we discussed simplifications in the dependence on the dimensional
regularization parameter $\eps$. 
In the case where the answer is given by iterated integrals, we discussed
a particularly simple canonical form of the equations and various aspects of their solution.

We discussed two systematic strategies
for obtaining this form, the first one being algebraic in nature, using algorithmic ideas available in the mathematical literature, and the second, more geometric one, using properties of the original loop integrand.

Although some of the algorithmic ideas of the first approach were already used in some form
or another in the physics literature, we hope that this presentation clarifies
in particular which aspects of the proposal of \cite{Henn:2013pwa} are always true, 
and which are conjectural. In particular, we explained how elliptic and more
complicated functions appear from this point of view.

The second approach is based on choosing loop integrals with simple
generalized cuts (and in particular leading singularities), as 
proposed in \cite{ArkaniHamed:2010gh}, and 
applied to the DE method in \cite{Henn:2013pwa}.
Although unitarity cuts are frequently used in other contexts, e.g. for finding coefficients of
loop integrals in unitarity-based calculations, see e.g. \cite{Bern:1994zx,Johansson:2012sf,Mastrolia:2012an,Badger:2013gxa} and 
references therein, to our surprise \cite{ArkaniHamed:2010gh} does not seem
to be widely known in the QCD community, and we hope that these lecture
notes help to disseminate these useful ideas.

In practice, we have found that a combination of the two methods is most 
promising: after choosing a basis according to the unitarity cut analysis,
the DE are usually either already in the canonical form, or very close to it,
so that the remaining transformations are simple to find.

There are several interesting avenues for exploration, and related work we wish to mention.

We have mostly focused around expansion near $D=4-2 \eps$ dimensions.
The same ideas outlined here can also be applied for expansions around other values
of the dimension. In particular, it is easy to see from the Feynman representation that Feynman integrals in $D$ and $D\pm 2$ dimensions are related \cite{Tarasov:1996br}. 
Therefore, the existence of a uniform weight basis in $D=4-2\eps$ dimensions implies 
the existence of a similar basis in dimensions related by multiples of $2$. 
This fact should imply non-trivial constraints on the matrices appearing in the canonical
form (\ref{canonical1}) of the differential equations.
Similarly, one can apply this method to study Feynman integrals in $D=3-2\eps$ dimensions,
e.g. with applications in ABJM theory.

The dimensional shifts mentioned above can also be helpful in finding a uniform weight
basis. For example, the four-dimensional pentagon integral with a numerator discussed in section
\ref{section_leading_singularities} is equivalent to  scalar pentagon integral in six dimensions.
Likewise, the four-dimensional bubble integrals with a doubled propagator can be equivalently
understood as two-dimensional bubble integrals with standard propagators.

It is important to emphasize that despite the simple dependence on $\eps$, the
differential equations are valid for any dimension, not just for small $\eps$. Therefore,
they can also be used as a starting point for writing down a solution for general dimension.
In this case one obtains hypergeometric functions and their generalizations. 
It would be interesting to connect this to the dimensional recurrence method of refs. \cite{Tarasov:1996br,Tarasov:2000sf,Lee:2009dh}.
 
We wish to mention that in the case of  integrals having neither UV nor IR divergences,
additional simplifications occur when evaluating the differential equations directly
in $D=4$ dimensions. In fact, it is possible to set up the IBP technique directly for the
case of finite integrals. For an example and more ample discussion, see ref.  \cite{Caron-Huot:2014lda}.
 
We have briefly discussed how elliptic and more complicated functions appear in this
setup. The question how to treat such functions systematically is an important conceptual 
(see e.g. ref. \cite{2011arXiv1110.6917B})
and practical question, e.g. for scattering amplitudes involving top quarks.

The question whether a given class of Feynman integrals can be evaluated in terms
of iterated integrals is crucial to an approach based on direct integration in Feynman 
parameter space \cite{Bogner:2014mha,Panzer:2014caa}, where it is related to
the question of linear reducibility.
The differential equations method also applies to cases that are outside of this class. 

Given a Feynman integral/graph, it should be possible in principle to determine
what class of functions arises (and in the case of iterated integrals, what the alphabet is) 
from its propagator structure. See e.g. \cite{2009arXiv0910.0114B} for work in this direction.
The transcendental weight properties of certain Feynman graphs have also been studied in \cite{2011CMaPh.301..357B}. 

As we have discussed, it is straightforward to obtain series representations from the differential equations. It would be interesting to explore the relationship to representations in terms of nested sums, see e.g. \cite{2013JMP....54h2301A}.

The criteria for finding an optimal basis
as presented here had the main purpose of transforming the DE into a canonical
form, which in turn makes e.g. their singularity structure manifest and makes it obvious
which class of functions is needed for their solution.
The main tool for this is the analysis of generalized cuts and leading singularities,
as proposed in ref. \cite{Henn:2013pwa}. 
We already mentioned that leading singularities were also used to construct
loop integrands having desirable physical properties, such as good behavior
in the infrared \cite{ArkaniHamed:2010kv,Drummond:2010mb,ArkaniHamed:2010gh}.
Such properties are usually lost when employing IBP, since the
IBP relations mix for example IR and UV poles.
We wish to emphasize that the requirement of a certain IR or UV behavior
at the level of the loop integrand, i.e. disregarding IBP relations,
is in general a stronger one, and we expect this to be of use e.g. when 
constructing a natural basis for higher loop integrands.
This is closely connected to the generalized unitary approach
for computing coefficients of loop integrals in a certain basis, see e.g. \cite{Johansson:2012sf,Mastrolia:2012an,Badger:2013gxa}
and references therein.

\section*{Acknowledgment}
J.M.H. is supported in part by the DOE grant DE-SC0009988,
and by the Marvin L. Goldberger fund.
These lecture notes are based on lectures given in 2013 at the LMS Durham Symposium
``Polylogarithms as a Bridge between Number Theory and Particle Physics'',
but also include newer material presented at seminar talks given in 2014 at UCLA, LAPTH Annecy,
MIAPP, and at the lecture series given at the Nordita School on Integrability, 2014, Stockholm.
It is a pleasure to thank F.~Caola, S.~Caron-Huot, A.~Grozin, G.~Korchemsky, P.~Marquard, K.~Melnikov, A.~Smirnov and V.~Smirnov for collaboration on topics presented here, and R.~Lee for discussions.

\bibliographystyle{JHEP} 

\bibliography{delectures}

  \end{document}